\documentclass[journal]{IEEEtran}
\ifCLASSINFOpdf
  \usepackage[pdftex]{graphicx}
\else
  \usepackage[dvips]{graphicx}
\fi
\usepackage{amsmath}
\usepackage{caption}

\usepackage{algorithmic}
\usepackage[ruled,vlined]{algorithm2e}
\usepackage{comment}
\usepackage{enumitem}
\usepackage{pifont}
\usepackage{bbm}

\makeatletter
\def\thickhline{%
  \noalign{\ifnum0=`}\fi\hrule \@height \thickarrayrulewidth \futurelet
   \reserved@a\@xthickhline}
\def\@xthickhline{\ifx\reserved@a\thickhline
               \vskip\doublerulesep
               \vskip-\thickarrayrulewidth
             \fi
      \ifnum0=`{\fi}}
\makeatother

\newlength{\thickarrayrulewidth}
\usepackage{multirow}
\usepackage{xcolor,colortbl}
\usepackage{cite}
\usepackage[bottom]{footmisc}
\begin{document}
%
\title{Learning Phone Recognition from Unpaired Audio and Phone Sequences Based on Generative Adversarial Network}
%
%

%

\author{Da-rong Liu,
        Po-chun Hsu, 
        Yi-chen Chen, 
        Sung-feng Huang,
        Shun-po Chuang,
        Da-yi Wu,
        and~Hung-yi Lee,
\thanks{D.-R. Liu, P.-C. Hsu, S.-P. Chuang, Y.-C. Chen and S.-F. Huang are with the Graduate Institute of Communication Engineering, National Taiwan University, Taipei 10617, Taiwan (e-mail: f07942148@ntu.edu.tw; f07942095@ntu.edu.tw; f04942141@ntu.edu.tw; f06942069@ntu.edu.tw; f06942045@ntu.edu.tw).}
\thanks{D.-Y Wu is with National Taiwan University, Taipei 10617, Taiwan (e-mail: ericwudayi2@gmail.com).}
\thanks{H.-Y Lee is with the Department of Electrical Engineering, National Taiwan University, Taipei 10617, Taiwan (e-mail: hungyilee@ntu.edu.tw).}
\thanks{This work was supported in part by the Ministry of Science and Technology, R.O.C, under Contract 110-2628-E002-001 and 110-2223-E-002-007-MY3.}
}

%
%

\markboth{Journal of \LaTeX\ Class Files,~Vol.~14, No.~8, August~2015}%
{Shell \MakeLowercase{\textit{et al.}}: Bare Demo of IEEEtran.cls for IEEE Journals}
%



\maketitle

\begin{abstract}
ASR has been shown to achieve great performance recently. However, most of them rely on massive paired data, which is not feasible for low-resource languages worldwide. This paper investigates how to learn directly from unpaired phone sequences and speech utterances. We design a two-stage iterative framework. GAN training is adopted in the first stage to find the mapping relationship between unpaired speech and phone sequence. In the second stage, another HMM model is introduced to train from the generator's output, which boosts the performance and provides a better segmentation for the next iteration. In the experiment, we first investigate different choices of model designs. Then we compare the framework to different types of baselines: (i) supervised methods (ii) acoustic unit discovery based methods (iii) methods learning from unpaired data. Our framework performs consistently better than all acoustic unit discovery methods and previous methods learning from unpaired data based on the TIMIT dataset.

\end{abstract}


%
\IEEEpeerreviewmaketitle

\section{Introduction}
%
%
%
%
\IEEEPARstart{W}{ith} the rapid development of deep learning, automatic speech recognition (ASR) has achieved remarkable performance and has been widely used. 
However, the training of state-of-the-art ASR systems~\cite{chorowski2015attention, chiu2018state, gulati2020conformer, han2020contextnet} often relies on massive annotated data.
Therefore, for low-resource languages with scarce annotated data, sufficiently accurate speech recognition is difficult to achieve.
Compared to annotating audio data for low-resource, huge quantities of unannotated audio data are relatively easy to collect. 
If the machine can directly learn an ASR model from unannotated audio with unpaired text only, building an ASR system will be more feasible for low-resource languages. 
We call such a setting `unsupervised ASR.'

Unsupervised learning, while challenging, has been successfully applied on machine translation~\cite{artetxe2017unsupervised, conneau2017word, lample2017unsupervised} and performed comparably with supervised learning results.
Unsupervised neural machine translation often comprises two-step training. 
In the first step, a transformation from source to target language word embedding spaces could be learned in an unsupervised manner.
In these papers~\cite{artetxe2017unsupervised, conneau2017word, lample2017unsupervised}, generative adversarial network (GAN)~\cite{goodfellow2014generative, arjovsky2017wasserstein, gulrajani2017improved} is served as the core framework.
The GAN consists of a discriminator and a generator.
The generator takes the source language as input and outputs the target language. 
The discriminator learns to distinguish the real target language from the generator output, 
while the generator learns to `fool' the discriminator.
The generator and the discriminator will train iteratively.
In the end, the generator output will become more and more `similar to' the real target language.
After training, the word translation can be achieved by selecting the nearest target word embedding compared to the transformed source word embedding.
Then in the second step, a denoising sequence-to-sequence model is trained based on the word translation acquired in step one to get the final translation model. 
In both two steps, the model selection and hyperparameter tuning are based on the proposed unsupervised metrics instead of the paired development set to prevent using any supervision information.

The success in unsupervised neural machine translation led to our attempts on unsupervised ASR, since ASR is also a kind of translation, trying to learn the mapping relationship from speech space to text space.
As being the first step toward unsupervised ASR, we make three compromises (i) we conduct the research on the phone-level instead of text-level (ii) we have access to the number of phones in advance (iii) we have the development set with paired phone transcriptions to tune the hyperparameters, but not involved in the training process.
Specifically, we aim for the task of phone recognition, where the learning algorithm only accesses (i) unannotated speech utterances, (ii) unpaired phone sequences during training while still using the development set to select the hyperparameters.
For the rest of the article, we denote our setting as `unpaired' phone recognition to make it clear we focus on how to learn phone recognition from unpaired data.

This paper attempts to use the GAN framework on unpaired phone recognition, where the generator takes speech utterances as inputs and outputs phone sequences.
After iterative training between generator and discriminator, the generator will serve as the final phone recognition model.
However, there is some fundamental difference between phone recognition and machine translation. 
In unsupervised machine translation, we know that most discrete source tokens can be mapped to specific target tokens representing the same meaning, and this mapping can be achieved by the step one described in the second paragraph. 
However, in unpaired phone recognition, which learns the mapping from an utterance (a series of acoustic features) to a discrete sequence, we do not know which segment of acoustic features should be mapped to a phone.
It is because each phone in a speech utterance consists of a segment of consecutive frames of variable length.
However, the phone boundaries are often unknown in advance, which has made unpaired phone recognition difficult.


To address the segmentation problem, we include a phone segmentation module before the generator.
This module can segment the original utterance into a sequence of phone-level segments.
Then the generator maps each segment into a phone and outputs a phone sequence.
This phone segmentation module is performed in an unsupervised manner.
Many previous unsupervised phone segmentation methods~\cite{wang2017gate, michel2016blind, qiao2008unsupervised, kreuk2020self, franke2016phoneme, rasanen2014basic, kamper2020towards} can be used here.

After the GAN training, the generator serves as the (first-version) phone recognition model.
We propose to further boost the performance via `self re-training.'
Inputting unpaired speech to the generator, we can generate their corresponding `pseudo transcription'. 
Then we view the speech utterances and the pseudo transcriptions as paired data and train a Hidden Markov Model (HMM) in a supervised manner.
Although the pseudo transcriptions have some errors compared to the oracle transcriptions, the experiment shows that HMM, training from pseudo transcriptions, can still significantly boost the performance compared to the first-version model.
Moreover, we use the trained HMM to perform forced alignment on the speech utterances and obtain new segmentation.
This segmentation is more accurate than the result of the unsupervised phone segmentation method because this segmentation is from a certainly well-trained HMM (the experiment also supports this fact).
With the new, better segmentation, we can repeat the same procedure: GAN training, self re-training, and getting new segmentation.
The iteration will continue until the performance converges. 
With the proposed framework, we get 36.7\% phone error rate on TIMIT.
This result is comparable to the supervised method trained with 2.5\% to 5\% paired data.

\begin{figure*}
\centering
\includegraphics[width=\linewidth]{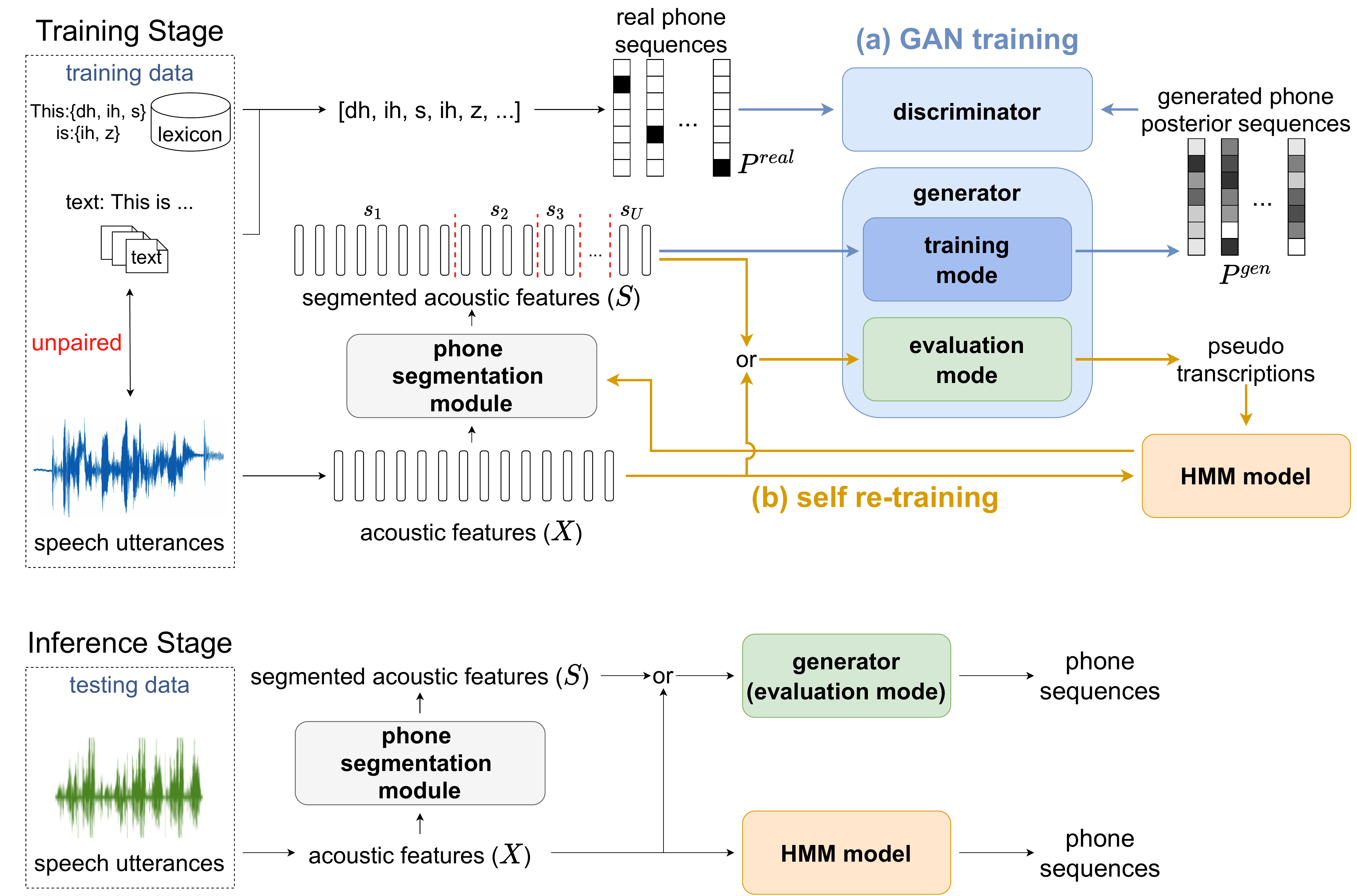}
\captionsetup{justification=centering,margin=2cm}
\caption{Framework overview: blue lines denote GAN training; yellow lines denote self re-training; 
black lines denote the paths which are not trainable. 
}
\label{fig:gan_framework}
\end{figure*}

The proposed two-stage iterative framework is summarised in figure~\ref{fig:gan_framework}. 
In the training stage, given a large collection of unpaired speech utterances and texts, speech utterances are segmented into phone-level segments by phone segmentation module, and texts are transformed into phone sequences by a lexicon.
In figure~\ref{fig:gan_framework}(a), GAN training is then conducted to map speech and phone sequence domains, which will be described in Section~\ref{gan training}.
In figure~\ref{fig:gan_framework}(b), self re-training is conducted after GAN training.
In this stage, another HMM is trained from the generator's output to boost the performance further.
HMM also provides more accurate segmentation information via forced alignment, which can be used for the next iteration of GAN training.
The details of self re-training will be described in Section~\ref{Refinement}.
In the inference stage, we can either use the generator or the HMM model as the final phone prediction model.
In this framework, the generator plays a vital role because it is used in GAN training, self re-training, and inference stage.
The generator is designed to have a specific output form under these different scenarios, and this will be discussed in Section~\ref{generator arch}.

\section{related work}
Learning phone recognition from unpaired data is related to unsupervised spoken term discovery (UTD) and acoustic unit discovery (AUD).
The main goal of UTD is to find clusters of speech segments that match their phonetic content, which may be syllables, words, or longer phrases. 
The existing methods for UTD can be broadly categorized into dynamic-time-warping-based (DTW-based) and embedding-based approaches.
The DTW-based approaches can be traced back to the segmental DTW algorithm by
\cite{park2006unsupervised}, followed by SWD-model by \cite{ten2007computational}, DP-ngram model by \cite{aimetti2009modelling}, to name a few.
In theory, DTW is very powerful since it enables pairwise alignment of any arbitrary sequences that consist of non-linear temporal distortions with respect to each other, and provides a measure of the degree of similarity. 
However, the complexity of DTW is $O(mnk)$ based on dynamic programming, where $m, n $ is the length of two sequences and $k$ is the dimension of the acoustic feature.
When $m, n $ is large, DTW will become computationally expensive in the inference time.
Besides, DTW approaches often use relatively primitive representation (e.g., MFCCs or posteriorgrams) and assume that spectral changes are time-synchronous, which may also affect the correctness.

Besides DTW, UTD can be done by embedding-based approaches.
These approaches attempt to capture the acoustic information in a speech segment of variable length by condensing it into a fixed-dimensional speech representation.
Speech segments containing the same contents are hopefully mapped to similar embeddings.
By pre-computing each possible speech segment into a vector, only dot product is needed when comparing the similarity of two speech segments in the inference time.
The complexity is $O(k)$, where $k$ is the embedding dimension.
Since the complexity of the embedding-based methods is independent of the segment lengths, they can compute much faster than DTW with long segments.
In the earlier work, embedding approaches were developed primarily in heuristic ways rather than learned from data. 
Graph-based embedding approaches are also used to represent audio segments as fixed-length vectors~\cite{levin2013fixed, levin2015segmental}. 
Recently, deep learning has been used to encode acoustic information as vectors~\cite{bengio2014word, chen2015query, kamper2016deep, he2016multi, settle2017query, maas2012word, chung2018speech2vec, holzenberger2018learning, kamper2019truly}. 
This transformation successfully produces vector spaces in which audio segments with similar phonetic structures are located in close proximity. 
By training a recurrent neural network (RNN) with an audio segment as the input and the corresponding word as the target, the outputs of the hidden layer at the last few time steps can be taken as the representation of the input segment~\cite{chen2015query, settle2016discriminative}. 
In~\cite{kamper2016deep}, the authors obtain embeddings by training a neural network that separates same-word and different-word pairs. 
In~\cite{chung2016audio}, a sequence-to-sequence auto-encoder training paradigm is proposed, which only uses word boundaries as supervision information.
It has been shown that the representation does contain phonetic information. 

On the other hand, AUD consists of discovering an inventory of phone-like discrete acoustic units from a set of untranscribed recordings. 
Nowadays, two major approaches are widely used in AUD: 
(i) neural-network-based models, which typically use an auto-encoder structure with a discretization layer~\cite{dunbar2019zero, dunbar2020zero, chen2020unsupervised, baevski2019vq, eloff2019unsupervised, chorowski2019unsupervised}.
(ii) non-parametric Bayesian generative-based models, which can be seen as infinite mixtures time series models~\cite{lee2012nonparametric,ondel2016variational, chen2017multilingual, ondel2019bayesian,yusuf2021hierarchical}.

Although all the above approaches can cluster speech segments according to phonetic structure and generate automatically discovered acoustic tokens, there is still a gap between mapping the discovered tokens into the human-defined tokens, for example phone and text. 
The lack of these mapping relationships has seriously limited the downstream applications because the core information of the speech signal is carried by the phones, and many downstream applications are only based on the transcriptions.
\cite{bansal2017towards} tries to transcribe some tokens into text while with the help of some extra annotations.
Our prior work \cite{liu2018completely} is the first work to successfully learn phone recognition with unpaired transcriptions by first clustering audio embeddings into a set of tokens and then using GAN to learn the mapping relationship between tokens and phones.
However, in \cite{liu2018completely}, the performance still relies on extremely good phone segmentation.
This was properly handled previously by a specially designed cost function called Segmental Empirical Output Distribution Matching (Segmental empirical-ODM)~\cite{liu2017unsupervised}, which considered both the n-gram probabilities across all output tokens and the frame-wise smoothness within each segment~\cite{yeh2018unsupervised}.
However, the loss term in \cite{liu2017unsupervised} includes an empirical average over all dataset inside a logarithmic function, which can be biased if we sample this empirical average by a mini-batch average. 
\cite{liu2017unsupervised} therefore proposes to use an extremely large batch size during training to reduce the biasing problem.
Besides, the n-gram probabilities considered here only include local statistics in the output sequences, while other information such as long-distance dependency is inevitably ignored. 
These problems are handled by another prior work ~\cite{chen2019completely}.
Based on \cite{liu2018completely}, \cite{chen2019completely} removes the quantization step and further proposes an iterative training algorithm that can correct the phone boundaries over the iterations and boost the performance. 
Compared to \cite{yeh2018unsupervised}, the discriminator in \cite{chen2019completely} considers all possible information from the dataset, not limited to the n-gram local statistics.
Besides, \cite{chen2019completely} works well with a reasonable batch size, which makes it feasible when the computational resource is limited.
\cite{chung2018unsupervised, chung2019towards} also use GANs to map the word-level speech embedding space~\cite{chung2018speech2vec} to word space, and achieve promising performance on spoken word classification, speech translation, and spoken word retrieval.
In \cite{chung2018unsupervised, chung2019towards}, the word boundaries are also generated automatically with speech only.

This journal paper is an extension of the previous conference papers~\cite{chen2019completely}. 
In this paper, we first propose a training framework, which includes GAN training and self re-training.
In the GAN training, \cite{chen2019completely} only describes one kind of architecture design of GAN.
While in this paper, all possible generator architectures are categorized into  segment-wise generator and frame-wise generator.
These two kinds of generators are discussed in detail and shown that the former performs better when we have oracle boundaries, while the latter performs better when we do not.
Different discriminator architectures are also discussed and compared to each other.
Besides, \cite{chen2019completely} has shown the effectiveness of self re-training.
In this paper, we further give the analysis and evidence that this improvement is brought from the improvement of phone segmentation over iterations.
Finally, we also analyze where the prediction errors happen, which has also not been done before.

\section{GAN training}\label{gan training}
In the training stage, we have a bunch of unpaired speech utterances that are represented as acoustic features $X$, and text.
We transform all text into phone sequences with the aid of a lexicon, and denote it as `real phone sequences' $P^{real}$.
We assume $P^{real}$ is representative and diverse enough to capture the distribution of all reasonable phone sequences. 
Our target is to learn a function $f: X\rightarrow P$, which maps speech to phone sequences, such that the distribution of $f(X)$ is close to the distribution of $P^{real}$.

The overview of GAN training is illustrated in figure~\ref{fig:gan_framework} (a).
The speech utterances are first represented as acoustic features and segmented into phone segments, which is described in Section \ref{phone segment}.
The segmented acoustic features are then inputted into the generator, and the generator outputs `generated phone posterior sequence.'
In Section~\ref{generator}, we only describe the input/output format of generator while leaving the architecture details to Section \ref{generator arch}.
In \ref{preprocess of real phn}, we talk about the preprocessing of real phone sequences $P^{real}$.
Then in Section \ref{discriminator}, we describe the discriminator details.
Finally, the optimization details of the GAN training process are discussed in Section \ref{gan training math}.

\subsection{Phone Segmentation Module}\label{phone segment}
The phone segmentation module segment the input acoustic feature sequence $X$ into a sequence of phone-level segments $S = \{s_1, ..., s_U\}$, where $s_i$ represents a segment of several consecutive acoustic features.
In the first iteration, because no supervised information is provided, this can only be done by unsupervised phone segmentation methods~\cite{wang2017gate, michel2016blind, qiao2008unsupervised, kreuk2020self, franke2016phoneme, rasanen2014basic, kamper2020towards}.
After the second iteration, we can perform forced alignment with the trained HMM model to generate the phone boundaries.

\subsection{Generator}\label{generator}
The generator takes a segmented acoustic feature sequence $ S \equiv \{s_1, ..., s_U\}$ as input, and outputs a phone distribution sequence $P^{gen}$, which we referred as `generated phone posterior sequence.'
Formally speaking, we denote,
\begin{align}
 & P^{gen} = \mathcal{G}(S)
 \\
 & P^{gen} \equiv (p^{gen}_1, p^{gen}_2, ... p^{gen}_U)
\end{align}
where $\mathcal{G}$ denotes the generator, $U$ is the length of the output phone sequence and $p^{gen}_i \in \mathcal{R}^{|S|}$ is the $i^{th}$ posterior over the phone set $\mathcal{S}$, which means:
\begin{align}
 & \sum^{|\mathcal{S}|}_{j=1} p^{gen}_i[j] = 1, \forall i \in [1,U]
\end{align}
where $p^{gen}_i[j]$ is the $j^{th}$ element of $p^{gen}_i$.
The design of the generator will be discussed in Section \ref{generator arch}.

\subsection{Preprocessing of Real Phone Sequence}\label{preprocess of real phn}
Corresponds to the generated phone posterior sequences, each real phone sequence $P^{real}$ is represented as a one-hot encoding sequence.
In this paper, we slightly abuse the notation $P^{real}$ to also refer to the phone sequence in one-hot encoding form.
Hence, we can mathematically denote:
\begin{align}
P^{real} \equiv (p^{real}_1, p^{real}_2, ... p^{real}_V)
\end{align}
where $V$ is the number of phones in the sequence and $p^{real}_i \in \mathcal{R}^{|\mathcal{S}|}$ is the one-hot encoding of the $i^{th}$ phone in the sequence.

\subsection{Discriminator}\label{discriminator}
The discriminator $\mathcal{D}$ is trained to distinguish between  $P^{gen}$ and $P^{real}$.
We follow the Wasserstein GAN (WGAN) framework~\cite{arjovsky2017wasserstein}.
The input of the discriminator is a sequence of phone posteriors, like $P^{gen}$ and $P^{real}$ ($P^{real}$ as one-hot encoding is also a kind of posterior); the output of the discriminator is a scalar.
The scalar is expected to be high when the input is $P^{real}$, while to be low when the input is $P^{gen}$.
Multiple discriminator architectures are feasible for this input/output form.
The detailed performance comparisons will be conducted in Section~\ref{exp: comparison dis arch}.

\subsection{Optimization Formulation of GAN Training}\label{gan training math}
\subsubsection{discriminator loss}
The loss for training the discriminator $\mathcal{D}$ follows the concept of WGAN \cite{arjovsky2017wasserstein, gulrajani2017improved} with gradient penalty~\cite{gulrajani2017improved}:
\begin{equation}
      \mathcal{L}_\mathcal{D} = 
      \frac{1}{K} \sum_{k=1}^{K}\mathcal{D}(\uppercase{p}^{gen(k)})   
       - \frac{1}{K} \sum_{k=1}^{K}\mathcal{D}(\uppercase{p}^{real(k)}) + \alpha \mathcal{L}_{gp},
  \label{discriminator_loss} 
\end{equation}
where $\mathcal{D}(P)$ is the scalar output of the discriminator for an input sequence $P$, $K$ is the number of training examples in a batch, and $k$ is the example index. $\alpha$ is the weight for the gradient penalty $\mathcal{L}_{gp}$:
\begin{equation}
      \mathcal{L}_{gp} = 
      \frac{1}{K} \sum_{k=1}^{K}( ||\nabla_{\uppercase{p}^{inter(k)}} \mathcal{D}( \uppercase{p}^{inter(k)})||-1 )^2,
  \label{gp_loss} 
\end{equation}
where $\uppercase{p}^{inter}$ is the interpolation of $\uppercase{p}^{real}$ and  $\uppercase{p}^{gen}$:
\begin{align}
 & \uppercase{p}^{inter} = \epsilon\uppercase{p}^{real} + (1-\epsilon)\uppercase{p}^{gen}
 \\
 & \epsilon \sim Uniform[0,1]
\end{align}
Because $\uppercase{p}^{real}$ and $\uppercase{p}^{gen}$ may have different sequence lengths, $\uppercase{p}^{real}$ and $\uppercase{p}^{gen}$ are first truncated to the shorter length of the two before calculating $\uppercase{p}^{inter}$.
This additional term has been proved useful in stabilizing the GAN training and preventing the gradient vanishing problem~\cite{gulrajani2017improved}.

\subsubsection{generator loss}
The generator loss is:
\begin{equation}
\mathcal{L}_{\mathcal{G}} = -\frac{1}{K} \sum_{k=1}^{K}\mathcal{D}(\uppercase{p}^{gen(k)}),
  \label{generator_loss} 
\end{equation}
The generator and the discriminator are learned to minimize the loss iteratively, so the generator can eventually map speech utterances to phone sequences `looking real.'


\section{Self Re-training}
\label{Refinement}
After the GAN is well trained, we decode the training set into phone sequences by the generator inside the GAN.
These GAN-generated phone sequences are taken as pseudo transcriptions for the training speech utterances to train a new phone HMM model, which can serve as a new phone recognition model.
Besides, the phone HMM can be used to perform forced alignment for the training set and get new segmentation boundaries, which are then used to start a new iteration of GAN training as described in Sections~\ref{gan training}, then another HMM is trained.
This training procedure can be performed iteratively until convergence.
In the inference stage, we can either use the generator or HMM model as the final phone recognition model.
The whole training/inference algorithm is depicted in Algorithm \ref{alg:algorithm}.

\begin{algorithm}
  \caption{Whole Framework} 
  \label{alg:algorithm}
  \SetKwBlock{kwtrain}{Training Stage:}{}
  \SetKwBlock{kwinf}{Inference Stage:}{}
  \kwtrain{
      \KwIn{Real phone sequences $\uppercase{p}^{real}$, Speech utterances $X$}
      \KwOut{generator, HMM}
      \SetKwBlock{kwinit}{Initialize:}{}
      \kwinit{unsupervised phone segmentation boundaries $b$}
      \SetKwBlock{kwgan}{GAN Training: }{end}
      \SetKwBlock{kwself}{Self Re-training: }{end}
      \While{not converged}{
        \kwgan{
            Given $b$, train GAN in an unsupervised manner\;
            Obtain transcriptions $T$ of each speech utterance using the generator within the GAN\;}
        \kwself{
            Given $T$, train the HMMs\;
            Obtain a new $b$ by forced alignment with the HMMs\;}
      } 
  }
  \kwinf{
      \KwIn{Speech utterances $X$, generator, HMM}
      \KwOut{predicted phone sequences}
      Obtain predicted phone sequences by the generator or HMM\;
  }
\end{algorithm}

\section{Generator}\label{generator arch}
There are two modes for the generator: training mode and evaluation mode.
In the training mode, which is used in the GAN training, the generator outputs generated phone posterior sequence $P^{gen} $, which is a sequence of phone posteriors.
In the evaluation mode, which is used in the self re-training and the inference stage, the generator outputs phone sequence, which is a sequence of discrete tokens.
In Sections~\ref{subsect:segment_generator} and ~\ref{subsect:frame_generator}, we categorize generators into two kinds of architectures: segment-wise generator, frame-wise generator.
Then their computation details in both training and evaluation mode are discussed.
In Section~\ref{subsubsect:softmax}, we further review Gumbel-Softmax, a special kind of softmax, to use in the training mode of generator.
Using Gumbel-Softmax is shown to achieve better performance in the experiments.


\begin{figure}
\centering
\includegraphics[width=\linewidth]{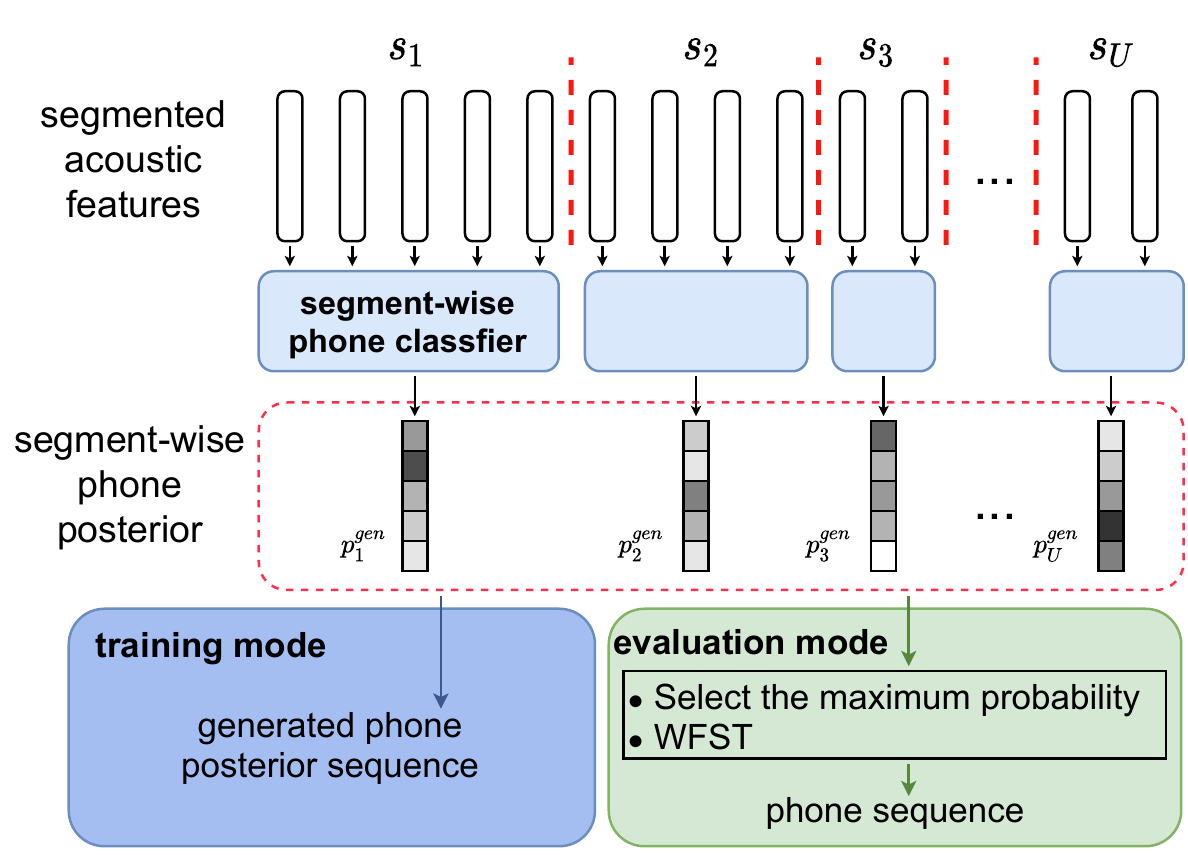}
\caption{Segment-wise generator
}
\label{fig:segment_generator_modules}
\end{figure}

\begin{figure}
\centering
\includegraphics[width=\linewidth]{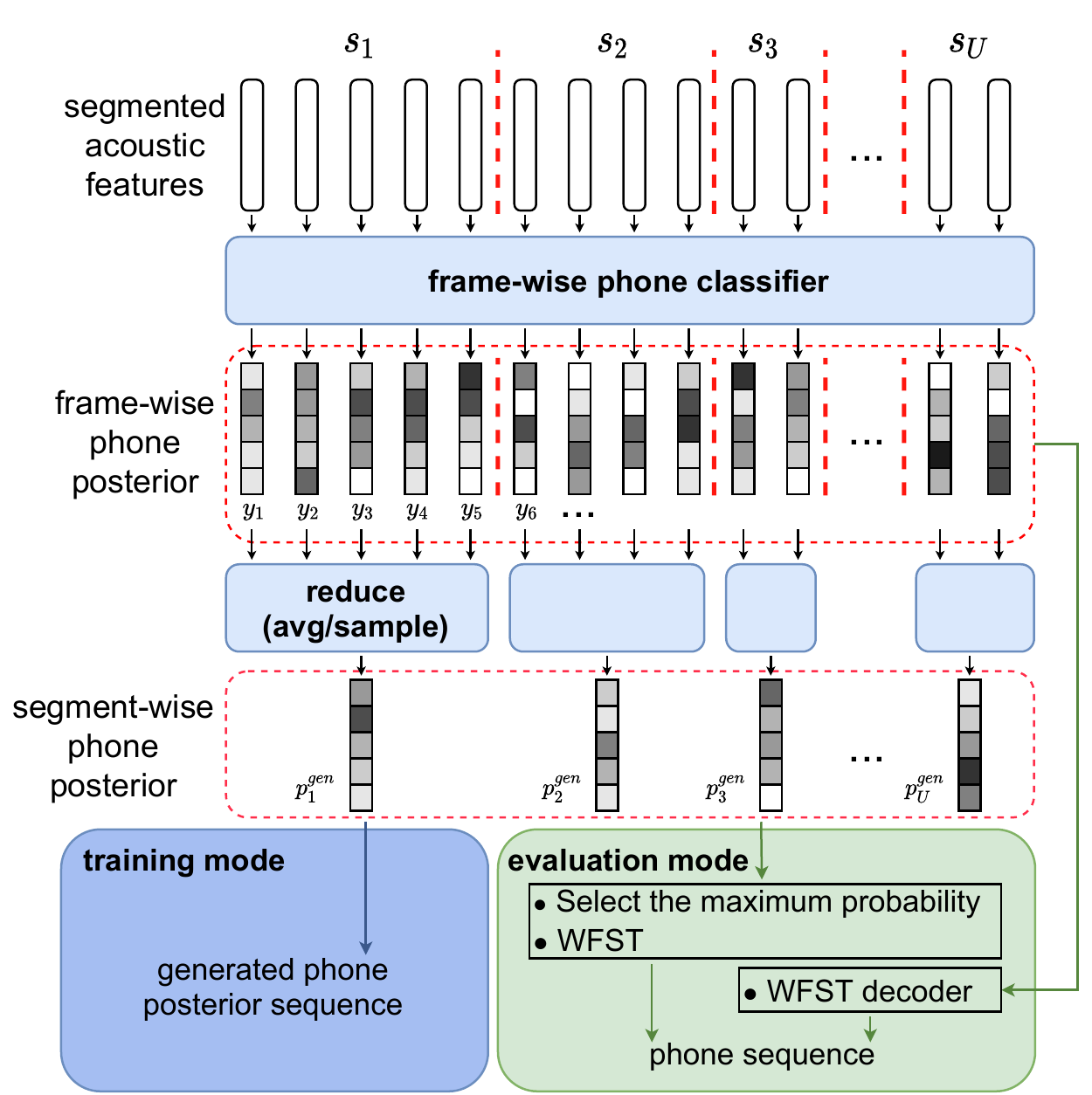}
\caption{Frame-wise generator}
\label{fig:frame_generator_modules}
\end{figure}

\subsection{Segment-wise Generator}\label{subsect:segment_generator}
Figure \ref{fig:segment_generator_modules} illustrates the training and evaluation mode of the segment-wise generator.
Each phone-level segment is first passed through a segment-wise phone classifier and outputs segment-wise phone posterior.
In the training mode, the segment-wise phone posterior is directly used as generated phone posterior sequence $P^{gen}$.
In the evaluation mode, we have two options to map the segment-wise phone posterior into phone sequence.
\subsubsection{Select the maximum probability}
The output phone sequence is generated by collecting the phone with the largest probability over all segment-wise phone posteriors.

\subsubsection{WFST}
Weighted Finite State Transducer (WFST) is a well-known speech decoder that can incorporate language model (LM) information and directly output the recognition result from a series of phone posteriors.
Instead of simply getting the phone with the largest probabilities of each posterior, 
we decode the phone sequence with the highest probability while considering phone n-grams.

\subsection{Frame-wise Generator}\label{subsect:frame_generator}
Figure \ref{fig:frame_generator_modules} illustrates the training and evaluation mode of the frame-wise generator.
Unlike segment-wise generator, in this architecture, frame-wise phone posteriors are first generated, which implies that phone boundaries are not used in the phone posterior generation process.
In the training mode, each segment of posterior is passed through a `reduce' process to reduce a segment of posteriors into one posterior.
In this paper, we introduce two types of reduce methods: average and sample.

\subsubsection{average}\label{subsubsect:average}
Each segment of frame-wise posteriors are averaged into a segment-wise posterior, which is denoted as:
\begin{align}
 & p^{gen}_i = \frac{1}{end_i-start_i+1}\sum_{j=start_i}^{end_i}y_j
\end{align}
where $start_i$ and $end_i$ denote the start and end index of the $i^{th}$ segment and $y_j$ is the $j^{th}$ frame posterior. 

\subsubsection{sample}\label{subsubsect:sample}
Instead of average a series of posteriors within a segment, we simply sample one of them at each training step.

When using frame-wise generator, to make the phone distribution inside a phone-level segment being consistent, we further introduce a loss, called `intra-segment loss' $\mathcal{L}_{intra}$:
\begin{equation}
\mathcal{L}_{intra} = -\frac{1}{K} \sum_{k=1}^{K}\sum_{i, j \in s_k}(y_i-y_j)^2,
  \label{intra_loss} 
\end{equation}
This loss serves as a regularization term for the generator loss described in equation~\ref{generator_loss}, so the new generator loss will be:
\begin{equation}
\mathcal{L}_{G} = -\frac{1}{K} \sum_{k=1}^{K}D(\uppercase{p}^{gen(k)})+\lambda \mathcal{L}_{intra},
  \label{new_generator_loss} 
\end{equation}
where $\lambda$ is a tunable weight.
To compute $\mathcal{L}_{intra}$, instead of exhaustively averaging over all permutations within a segment, we sample 10 pairs of frames in the implementation.

In the evaluation mode, we can either generate a phone sequence from segment-wise posterior or directly from frame-wise posterior.
If choosing from segment-wise posterior, then the case will be similar to the segment-wise generator.
To get the phone sequence, we can either select (1) select the maximum probability or (2) WFST, which are both described in Section~\ref{subsect:segment_generator}.
Only `average' method is used for the reduce module in the evaluation mode because the sampling process will bring randomness, which is not suitable for evaluation.
Another choice is to generate from frame-wise posterior.
In this case, we can also use the WFST decoder because the WFST decoder can also take posterior with repeated phone as input.
In this case, the segmentation boundaries are even no longer necessary (in the evaluation stage).

\subsection{Gumbel-Softmax}\label{subsubsect:softmax}
\begin{figure}
\centering
\includegraphics[width=\linewidth]{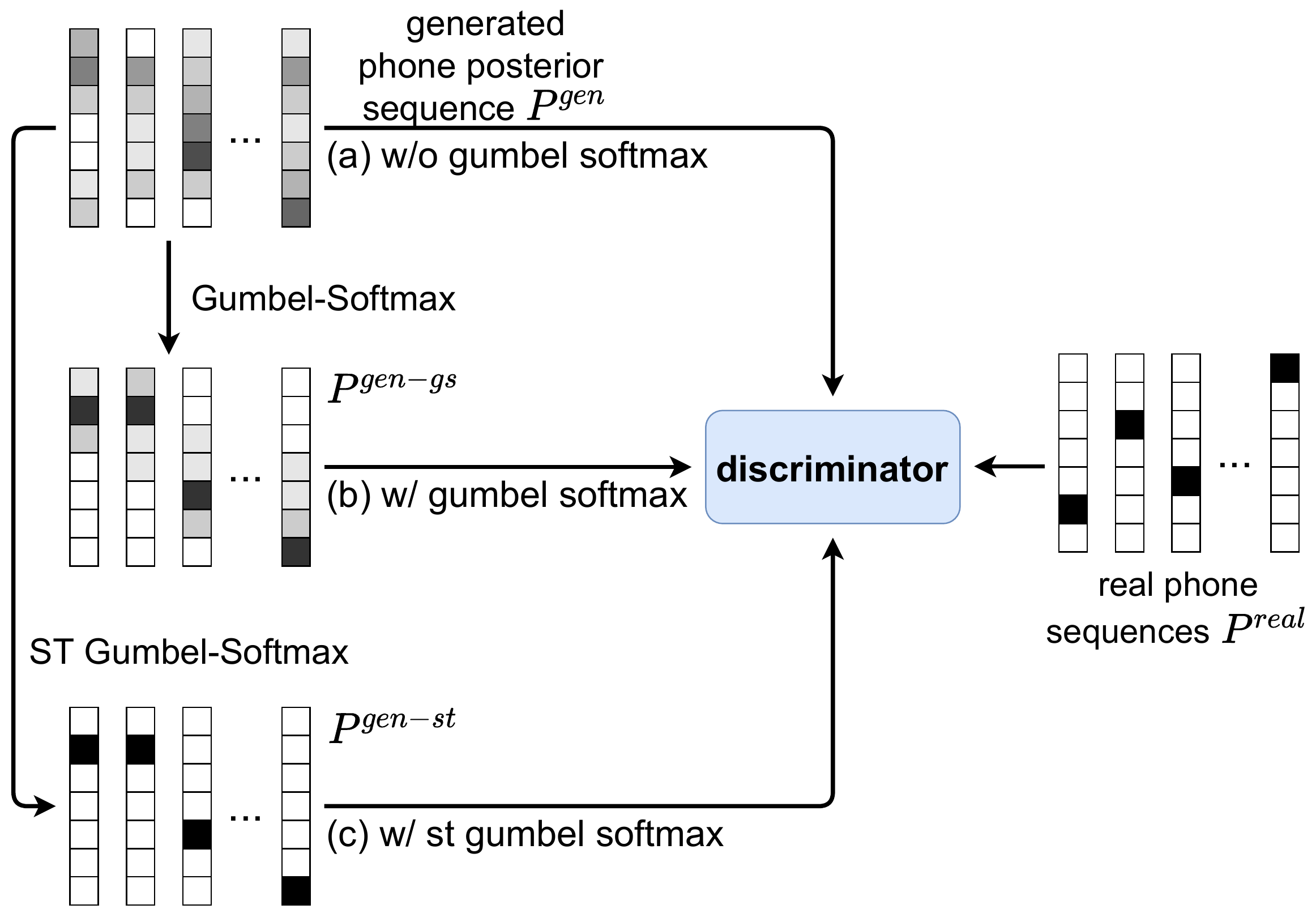} 
\caption{In Figure \ref{fig:gan_framework},  generated phone posterior sequence $P^{gen}$ is directly fed to the discriminator (path a) along with the real phone sequence $P^{real}$. 
Due to $P^{gen}$ is the output of softmax, which is a soft distribution, while $P^{real}$ is a one-hot distribution, the discriminator may distinguish two sequences easily~\cite{gulrajani2017improved}.
Although gradient penalty can somehow mitigate the problem, we further pass through Gumbel-Softmax~\cite{jang2016categorical, maddison2016concrete} to make $P^{gen}$ a little sharper (path b) or pass through straight-through Gumbel-Softmax~\cite{jang2016categorical, maddison2016concrete} to make $P^{gen}$ also be a one-hot distribution (path c) and see if the performance improves.} 
\label{fig:gumbel}
\end{figure}

In the training mode, we have an option to add `Gumbel-Softmax' module at the output of generated phone posterior sequence $P^{gen}$ before passing to the discriminator, as illustrated in Figure \ref{fig:gumbel}.
Gumbel-Softmax mimics the categorical sampling procedure while making the sampling process differentiable, and is only used in the training mode and not involved in the evaluation mode. 
Three scenarios are considered in the experiments.
\subsubsection{Without Gumbel-Softmax} The generated phone posterior sequence $P^{gen}$ is directly fed into discriminator.
\subsubsection{With Gumbel-Softmax}
We replace the softmax layer at the output of $P^{gen}$
\begin{equation}
p^{gen}_{i}[j] = \frac{\exp(l_{ij})}{\sum_{j=1}^{|\mathcal{S}|}\exp(l_{ij}))},
\end{equation}
into
\begin{equation}
p^{gen-gs}_{i}[j] = \frac{\exp((l_{ij}+g_{ij})/\tau)}{\sum_{j=1}^{|\mathcal{S}|}\exp((l_{ij}+g_{ij})/\tau)},
\label{eq:gumbel}
\end{equation}
where $l_{i1}, ...l_{i|\mathcal{S}|}$ are the logits before softmax, $g_{i1}, ...g_{i|\mathcal{S}|}$ are i.i.d samples drawn from gumbel distribution $Gumbel(0, 1)$ and $\tau$ is the temperature.
Samples from Gumbel distribution $g_{ij}$ can be transformed from samples from uniform distribution:
\begin{align}
 & u_{ij} \sim Uniform(0,1)
 \\
 & g_{ij} = -\log(-\log(u_{ij}))
\end{align}
\subsubsection{With straight-through(ST) Gumbel-Softmax}
After passing through Gumbel-Softmax module, $p^{gen-gs}_{i}$ is sharper while not being an one-hot vector.
We can further convert the $p^{gen-gs}_{i}$ into one-hot encoding by taking argmax class from vector:
\begin{equation}
p^{gen-st}_{i} = onehot(\mathop{\arg\max}_{j} \ p^{gen-gs}_{i}[j]).
\label{eq:st_gumbel}
\end{equation}
According to Gumbel-Max trick~\cite{gumbel1954statistical, maddison2016concrete}, $p^{gen-st}_{i}$ is exactly a categorical sample sampling from the posterior $p^{gen}_{i}$~\cite{gumbel1954statistical, jang2016categorical}.
In the ST Gumbel-Softmax, because $onehot+argmax$ is a flat function, with zero gradients at most points and even non-differentiable at some points, we approximate the gradient:
\begin{equation}
\frac{\partial p^{gen-st}_{i}}{\partial p^{gen-gs}_{i}\hfill} \approx \mathbbm{1}.
\label{eq:approximate}
\end{equation}
in the backpropagation~\cite{jang2016categorical}.
Therefore, in the training of ST Gumbel-Softmax, we only calculate an approximation of the gradients, not the exact value.

\section{Experimental Setup}
\subsection{Dataset}
We use the TIMIT corpus to evaluate the performance.
TIMIT contains recordings of phonetically-balanced utterances.
Each utterance includes manually aligned phonetic/word transcriptions, as well as a 16-bit, 16kHz waveform file. 
For each utterance, 39-dim MFCCs are extracted with utterance-wise cepstral mean and variance normalization (CMVN) applied.
All models are trained based on the 48 phone classes.
We evaluate phone error rate (PER) on the standard test set, which contains 192 utterances, for 39 phone classes mapped from the 48 output classes of the model.

\subsection{Training Settings}
There are four settings for the following experiments, which are the permutations of \emph{match/nonmatch} and \emph{orc/uns}.
Each of them is respectively described below
\begin{enumerate}[]
\item \emph{match:} 
The utterances and the real phone sequences are matched but not aligned during training.
\item \emph{nonmatch:} 
There is no overlapping utterance ID between the utterances and the real phone sequences. 
Because TIMIT has different utterances recorded from the same sentences, there are still recordings of the same content (word sequence) in two sets. 
\item \emph{orc:} the oracle boundaries provided by TIMIT are used.
\item \emph{uns:} 
the initial boundaries are obtained automatically with GAS~\cite{wang2017gate}. 
\end{enumerate}   

\section{Architecture Discussion of GAN training}\label{exp:gan training}
We first discuss the performance of using different architecture designs in the GAN training.
Following~\cite{chen2019completely}, we select 4000 utterances in the original training set for training and others for validation. 
Under \emph{match} setting, all 4000 utterances are used as real phone sequences.
Under \emph{nonmatch} setting, 3000 utterances are taken as speech utterances while the phone transcriptions of the other 1000 utterances are taken as the real phone sequences.
In this section, we focus on the GAN architecture. 
Therefore, all experiments are conducted with training one iteration and without self re-training.

\subsection{Comparing Segment-wise and Frame-wise Generator}
\label{sebsec:seg_frame}
We first compare the performance of the segment-wise and frame-wise generators, which are described respectively in Section \ref{subsect:segment_generator} and Section \ref{subsect:frame_generator}. 
The Gumbel-Softmax is used in all training, and the temperature is set to 0.9.
The discriminator in both settings is a two-layer convolutional network. 
The first layer is a 1-D convolution bank~\cite{wang2017tacotron, lee2017fully} with kernel sizes 3,5,7,9, and channel size 256. 
The second layer is a convolution layer with kernel size 3 and channel size 1024. 
The gradient penalty weight $\alpha$ is set to 10.
In the \emph{uns} setting, we randomly remove 4\% of the phones and duplicate 11\% of the phones in the real phone sequences to generate augmented phone sequences to be used as real phone sequences.
We use RAdam~\cite{liu2019variance} optimizer for both generator and discriminator.
The batch size is set to 100.
When using the WFST decoder, we use unaugmented real phone sequences to train a 5-gram LM.
The AM/LM ratio and the self-loop probability are tuned according to the PER of the development set.
The specific GAN setting for each method are listed below: 
\subsubsection{settings of segment-wise generator}
In the segment-wise generator, we first use an LSTM with 512 hidden units to encode each segment into a fix-dimension vector.
Then we use a linear layer to project the vector into the phone posterior.
The learning rate for the generator and discriminator is set to 1e-2 and 1e-3, respectively, and the discriminator updates three times per generator update.
In the evaluation stage, we report the PER of 1. selecting the maximum probability from segment-wise phone posterior (\emph{Max Prob}) 2. using WFST on top of segment-wise phone posterior (\emph{WFST}).

\subsubsection{settings of frame-wise generator}
For the frame-wise generator, we take the concatenation of 11 windowed frames of MFCCs as the input feature.
The phone classifier is a one-layer DNN with 512 ReLU units with output classes equal to phone numbers.
Intra-loss weight $\lambda$ is set to 0.5. 
The learning rate for the generator and discriminator is set to 1e-3 and 2e-3, respectively, and the discriminator updates three times per generator update.
In the evaluation stage, we report the PER of 1. \emph{Max Prob} 2. \emph{WFST} 3. using WFST on top of frame-wise phone posterior (\emph{WFST\textsuperscript{*}}).

\begin{table*}[]
\caption{Performance of segment-wise and frame-wise generators: \emph{Max Prob} denotes select the phone with the highest probability of each segment-wise phone posterior. 
\emph{WFST} means using WFST decoder on top of segment-wise phone posterior; \emph{WFST\textsuperscript{*}} means using WFST decoder on top of frame-wise phone posterior.
\emph{Evaluation metric: PER in [\%].}}
\centering
\begin{tabular}{lllcccc}
\thickhline
\multicolumn{2}{c}{\multirow{2}{*}{}} & \multicolumn{1}{c}{\multirow{2}{*}{}} & \multicolumn{2}{c}{\emph{orc}} & \multicolumn{2}{c}{\emph{uns}}\\ \cline{4-7} 
\multicolumn{2}{c}{} & \multicolumn{1}{c}{} & \begin{tabular}[c]{@{}c@{}}\emph{match}\end{tabular} & \begin{tabular}[c]{@{}c@{}}\emph{nonmatch}\end{tabular} & \begin{tabular}[c]{@{}c@{}}\emph{match}\end{tabular} & \begin{tabular}[c]{@{}c@{}}\emph{nonmatch}\end{tabular} \\ \hline
\multicolumn{2}{l|}{\multirow{2}{*}{segment-wise generator}} & (a) \emph{Max Prob} & 23.07 & 35.83 & 68.66 & 82.73 \\
\multicolumn{2}{l|}{} & (b) \emph{WFST} & \bf{22.35} & \bf{33.89} & 67.10 & 82.44 \\ \hline \hline
\multicolumn{1}{l|}{\multirow{6}{*}{frame-wise generator}} & \multicolumn{1}{l|}{\multirow{3}{*}{average}} & (c) \emph{Max Prob} & 26.91 & 36.12 & 66.16 & 69.35 \\
\multicolumn{1}{l|}{} & \multicolumn{1}{l|}{} & (d) \emph{WFST} & 24.24 & 34.10 & 65.18 & 65.82 \\
\multicolumn{1}{l|}{} & \multicolumn{1}{l|}{} & (e) \emph{WFST\textsuperscript{*}} & 29.01 & 38.81 & 58.95 & 62.13 \\ \cline{2-7} 
\multicolumn{1}{l|}{} & \multicolumn{1}{l|}{\multirow{3}{*}{sample}} & (f) \emph{Max Prob} & 28.63 & 36.97 & 64.02 & 63.33 \\
\multicolumn{1}{l|}{} & \multicolumn{1}{l|}{} & (g) \emph{WFST} & 26.38 & 35.34 & 60.80 & 61.83 \\
\multicolumn{1}{l|}{} & \multicolumn{1}{l|}{} & (h) \emph{WFST\textsuperscript{*}} & 31.52 & 40.12 & \bf{55.70} & \bf{57.57} \\ 
\thickhline
\end{tabular}
\label{Tab:compare}
\end{table*}

Our results are shown in Table \ref{Tab:compare}.
First, \emph{WFST} performs consistently better than \emph{Max Prob} ((b) vs. (a), (d) vs. (c), (g) vs. (f)), which means incorporating LM information helps improve the PER.
In the evaluation stage, \emph{WFST\textsuperscript{*}} ((e)(h)) does not use boundary information. 
On the other hand, \emph{WFST} and \emph{Max Prob} are generated from segment-wise phone posteriors, thus using boundary information during evaluation. 
We observe that under \emph{orc} setting, where the boundaries are perfect, \emph{WFST} and \emph{Max Prob} performs better than \emph{WFST\textsuperscript{*}} ((c)(d) vs. (e), (f)(g) vs. (h), (a)(b) vs. (e)(h)).
Among them, the segment-wise generator performs the best ((a)(b) vs. others).
On the other hand, under \emph{uns} setting, where the boundaries are not accurate, \emph{WFST\textsuperscript{*}}, which does not use boundary information, performs better ((e)(h) vs. others).

When using frame-wise generator, using `average' as reduce module performs better than `sample' under \emph{orc} setting, while worse under \emph{uns} setting ((e) vs. (h)).
\begin{figure}
\centering
\includegraphics[width=\linewidth]{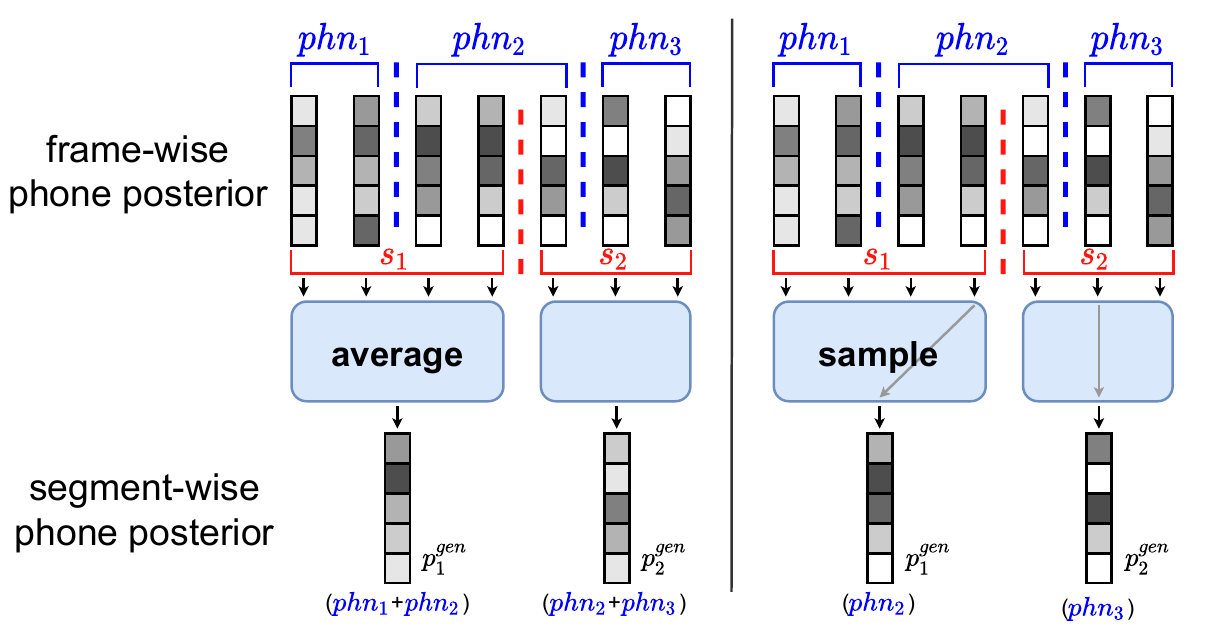}
\caption{
The blue lines denote the oracle phone boundaries; the red dashed lines denote the unsupervised generated phone boundaries. 
Here take frame-wise generator as an example and use \emph{uns} setting.
When using `average' as the reduce module, the generated segment-wise phone posterior may use the acoustic information from nearby phones.
For example, in the left part of the figure, $p^{gen}_2$ contains the acoustic information from $phn_2$ and $phn_3$.
On the other hand, this will not happen when using `sample' as the  reduce module.
For example, in the right part of the figure, each posterior only contains the acoustic information from one phone.
}
\label{fig:cross_segmentation}
\end{figure}
We hypothesize that during training and under the \emph{uns} setting, using `sample' is less susceptible to errors in segmentation because it will not utilize the acoustic information from nearby phone segments, while using `average' will.
This is further explained in figure~\ref{fig:cross_segmentation}.
For the following experiments, we keep using the frame-wise generator with `sample' reduce module attached with a frame-wise WFST decoder (\emph{WFST\textsuperscript{*}}), which has the best performance under \emph{uns} setting.

\subsection{Discussion of the Capacity of the Frame-wise Generator} \label{exp: gen_capacity}
The previous section uses one-layer DNN with 512 units, which is a relatively simple model, as the frame-wise generator.
Generally speaking, powerful models usually achieve better performances in supervised learning if not overfitting.
While in the GAN training, if the generator is powerful enough, the generator in principle can completely output a phone sequence that is unrelated to the input acoustic feature and converge to a bad result in the end.
In Table~\ref{tab:generator}, we compare two different generators: DNN and LSTM.
Both models use 512 hidden units.
Although LSTM has a larger model capacity, the experiments show that LSTM only achieves similar performance to DNN in \emph{orc} setting while worse in \emph{uns} setting.
This result supports the fact that using powerful generators is not always better in GAN training.

\begin{table}[]
\centering
\caption{Comparison between different frame-wise generators.
\emph{Evaluation metric: PER in [\%].}
}
\begin{tabular}{lcccc}
\thickhline
 & \multicolumn{2}{c}{\emph{orc}} & \multicolumn{2}{c}{\emph{uns}} \\ \cline{2-5}
 & \begin{tabular}[c]{@{}c@{}}\emph{match} \end{tabular} & \begin{tabular}[c]{@{}c@{}}\emph{nonmatch}\end{tabular} & \begin{tabular}[c]{@{}c@{}}\emph{match}\end{tabular} & \begin{tabular}[c]{@{}c@{}}\emph{nonmatch}\end{tabular} \\ \hline
(a) DNN  & 31.52 & \bf{40.12} & \bf{55.70} & \bf{57.57} \\ 
(b) LSTM & \bf{31.36} & 41.89 & 74.14 & 75.63 \\ 
\thickhline
\end{tabular}
\label{tab:generator}
\end{table}

\subsection{Using Gumbel-Softmax in Generator}
Then we investigate the effectiveness of using Gumbel-Softmax at the output of the generator in the training mode.
The temperature is set to 0.9.
The results are shown in Table~\ref{tab:softmax}.
\begin{table}[]
\centering
\caption{Performance of using Gumbel-Softmax. 
`ST Gumbel-S.' stands for straight through Gumbel-Softmax.
\emph{Evaluation metric: PER in [\%].}
}
\begin{tabular}{lcccc}
\thickhline
 & \multicolumn{2}{c}{\emph{orc}} & \multicolumn{2}{c}{\emph{uns}} \\ \cline{2-5} 
 & \begin{tabular}[c]{@{}c@{}}\emph{match} \end{tabular} & \begin{tabular}[c]{@{}c@{}}\emph{nonmatch} \end{tabular} & \begin{tabular}[c]{@{}c@{}}\emph{match} \end{tabular} & \begin{tabular}[c]{@{}c@{}}\emph{nonmatch} \end{tabular} \\ \hline
(a) w/o Gumbel-S. & 32.03 & 46.10 & 56.52 & 59.41 \\
(b) w/  Gumbel-S. & \bf{31.52} & \bf{40.12} & \bf{55.70} & \bf{57.57} \\
(c) w/ ST Gumbel-S. & 35.12 & 52.21 & 57.41 & 67.63 \\
\thickhline
\end{tabular}
\label{tab:softmax}
\end{table}
In all cases, using Gumbel-Softmax performs consistently better than w/o using Gumbel-Softmax ((b) vs. (a)), which validates the effectiveness of making the generator output sharper.
However, using ST Gumbel-Softmax performs worse than w/o using Gumbel-Softmax ((c) vs. (a)). 
This is counter-intuitive because ST Gumbel-Softmax directly converts the generator output into one-hot encoding, which is the sharpest possible output but performs the worst in the end.
We hypothesize this is because the gradient of ST Gumbel-Softmax is being approximated that makes the model converge to a bad point.

During evaluation, we also find that the frame-wise phone posterior will converge to different degrees of sharpness after training with different kinds of Gumbel-Softmax.
Table~\ref{tab:entropy} shows the average entropy of each posterior in the frame-wise phone posterior in the evaluation mode.
\begin{table}[]
\centering
\caption{Average entropy of each posterior in frame-wise phone posterior after training with different Gumbel-Softmax under \emph{orc}/\emph{match} setting.}
\begin{tabular}{lcc}
\thickhline
 & \multicolumn{1}{c}{\emph{train-set}} & \multicolumn{1}{c}{\emph{test-set}} \\ \hline
uniform distribution & \multicolumn{1}{c}{3.87} & \multicolumn{1}{c}{3.87} \\ \hline
(a) w/o Gumbel-Softmax & 0.10 & 0.11 \\ 
(b) w/ Gumbel-Softmax & 0.08 & 0.09 \\ 
(c) w/ st Gumbel-Softmax & 0.56 & 0.59 \\ 
\thickhline
\end{tabular}
\label{tab:entropy}
\end{table}
After GAN training, the frame-wise phone posterior converges to have the highest entropy using ST Gumbel-Softmax.
There is a large gap between ST Gumbel-Softmax and any other method.
We hypothesize that the model output before ST Gumbel-Softmax no longer has to be a sharper posterior to fool the discriminator because the output will be converted into a one-hot distribution $P^{gen-st}$ during training. 

\subsection{Comparing Discriminator Architecture} \label{exp: comparison dis arch}
In this section, we investigate how well the GAN performs under different discriminator architectures.
Four discriminator architectures are discussed and listed below\footnote{We do not include the LSTM-based discriminator.
Because the framework is based on WGAN~\cite{arjovsky2017wasserstein}, we need to calculate second-order derivatives during backpropagation, while the second differentiation of the LSTM module is currently not supported in Pytorch.}.
\subsubsection{transformer} Transformer~\cite{vaswani2017attention} comprises positional encoding and stacks of multi-head attention layer and layer normalization.
We use the transformer with 128 hidden units, 8 heads for multi-head attention, and 6 stacks in total.
\subsubsection{Conv-bank} Conv-bank~\cite{wang2017tacotron, lee2017fully} is the concatenation of a bank of 1-D convolutional filters with different kernel sizes. 
Conv-bank has been shown effective in modeling the input data with multi-resolution properties, which is suitable for speech data.
Speech has multi-resolution properties because speech consists of words, and words consist of different lengths of phones.
We use one layer conv-bank with kernel sizes 3, 5, 7, 9 to capture the phone patterns with different resolutions.
Hidden units are set to 256.
\subsubsection{Conv-bank+} Conv-bank is designed to capture different phone patterns, which are possible words.
We want to further model the interaction between these words.
The interaction of words can be viewed as a kind of language model.
Therefore, we stack one more convolutional layer with kernel size 3 and channel size 1024 to capture the relationship between neighbor words (phonetic pattern).
\subsubsection{Deep CNN}
We also use deep convolutional networks as the discriminator.
To compared with conv-bank based discriminator fairly, we set kernel size to 3 and 8 layers. 
In this setting, the receptive field will be 17, which is the same as conv-bank+.

The results are shown in Table~\ref{tab:discriminator}.
\begin{table}[]
\centering
\caption{Performance of using different discriminators. \emph{Evaluation metric: PER in [\%].}}
\begin{tabular}{lcccc}
\thickhline
 & \multicolumn{2}{c}{\emph{orc}} & \multicolumn{2}{c}{\emph{uns}} \\ \cline{2-5}
 & \begin{tabular}[c]{@{}c@{}}\emph{match} \end{tabular} & \begin{tabular}[c]{@{}c@{}}\emph{nonmatch}\end{tabular} & \begin{tabular}[c]{@{}c@{}}\emph{match}\end{tabular} & \begin{tabular}[c]{@{}c@{}}\emph{nonmatch}\end{tabular} \\ \hline
(a) transformer & 78.01 & 79.78 & 85.31 & 82.25 \\ 
(b) conv-bank & 32.23 & 41.77 & 57.24 & 61.74 \\ 
(c) conv-bank+ & \bf{31.52} & \bf{40.12} & \bf{55.70} & \bf{57.57} \\ 
(d) deep CNN & 53.89 & 56.50 & 69.83 & 68.19 \\ 
\thickhline
\end{tabular}
\label{tab:discriminator}
\end{table}
Although the transformer and deep CNN model are considered to be more expressive models compared to conv-bank, the performance is worse ((a)(d) vs. (b)(c)).
We hypothesize that under the condition of using a simple generator (following the results of Section~\ref{exp: gen_capacity}), the discriminator will distinguish `real phone sequence' and `generated phone posterior sequence' easily when it is too powerful, and can not correctly guide the generator~\cite{gulrajani2017improved}.
Finally, the conv-bank+ performs consistently a little better than conv-bank ((c) vs. (b)) and performs the best among all.

\subsection{Error Analysis}

\begin{figure}
\centering
\includegraphics[width=\linewidth]{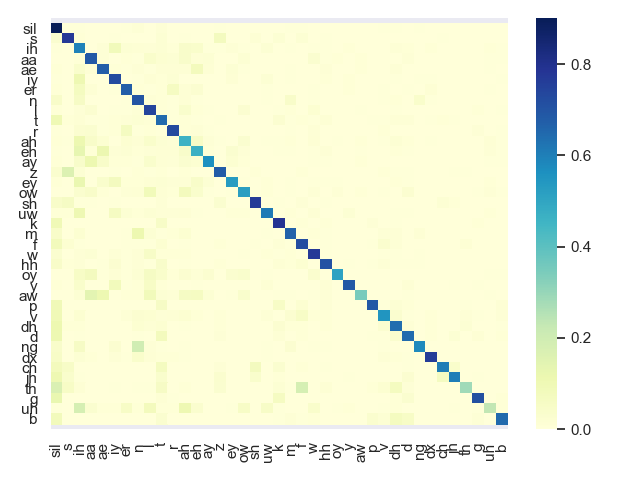}
\caption{The vertical axis represents the oracle phones of the frames; the horizontal axis represents the averaged frame-wise phone posterior. The phones are ordered according to the phone frequency. }
\label{fig:heatmap}
\end{figure}

Under \emph{orc/match}, we also visualize the heat map of the averaged frame-wise phone posterior to the corresponding oracle phone, which is illustrated in Figure~\ref{fig:heatmap}.
We observe that although the prediction of the high-frequency phones is a little more accurate than the low-frequency phones, the low-frequency phones still keep a certain degree of accuracy.
This means our method is robust to all phones, not only focus on the high-frequency phones.
In Table~\ref{tab:error_eg}, we also list the top 6 prediction error phone pairs.
\begin{table}[]
\centering
\caption{Top prediction error phone pairs}
\begin{tabular}{cccc}
\thickhline
rank & oracle phone & predicted phone & error percentage (\%) \\ \hline
\#1 & ng & n & 19.9 \\
\#2 & uh & ih & 18.5 \\
\#3 & th & f & 18.4 \\
\#4 & th & sil & 17.0 \\
\#5 & z & s & 16.2 \\
\#6 & aw & aa & 14.3 \\
\thickhline
\end{tabular}
\label{tab:error_eg}
\end{table}
We observe that the error mostly happens between two phones whose   pronunciations are similar, for example, \emph{n} and \emph{ng}, \emph{z} and \emph{s}, \emph{aw} and \emph{aa}.

\section{Compared to previous works}\label{exp:compare_previous}
In this section, we compare our framework to previous methods.
To compare to previous works fairly, we follow the standard TIMIT split and the setting in~\cite{yeh2018unsupervised}.
Under \emph{match} setting, all 3696 utterances are used as real phone sequences.
Under \emph{nonmatch} setting, we follow~\cite{yeh2018unsupervised} to split the train and dev set into 3000 and 1096 to serve as utterances and real phone sequences\footnote{Because \cite{yeh2018unsupervised} does not release the nonmatch split, we follow the split number while the actual split may be different.}.
During GAN training, $\alpha$ is set to 18, and the frame-wise generator is a 256-256 DNN.
Other training details are the same as Section~\ref{sebsec:seg_frame}.
In self re-training, the HMM (monophone and triphone) training followed the standard recipes of Kaldi~\cite{Povey_ASRU2011}. 
Linear Discriminant Analysis (LDA) and Maximum Likelihood Linear Transform (MLLT) are applied to MFCCs for model training. 
While decoding, the setting of LM is the same as the WFST decoder used after GAN training.
The framework runs for one iteration under \emph{orc} setting, four iterations under \emph{uns} setting.
In the \emph{uns} setting, we augment the real phone sequence by randomly removing 4\% and duplicating 11\% of phones in the first iteration while do not use augmentation after the second iteration. 

\begin{table}[]
\centering
\caption{Comparison of different methods.
\emph{Evaluation metric: PER in [\%].}
}
\begin{tabular}{lcccc}
\thickhline
\multicolumn{5}{c}{(I) Supervised} \\ \hline
\multicolumn{1}{l}{(a) RNN Transducer~\cite{yeh2018unsupervised}} & \multicolumn{4}{c}{17.70} \\
\multicolumn{1}{l}{(b) standard HMMs} & \multicolumn{4}{c}{21.60} \\
\multicolumn{1}{l}{(c) \emph{supervised generator}} & \multicolumn{4}{c}{23.24} \\
\multicolumn{1}{l}{(d) FS-RNN~\cite{ratajczak2017frame}} & \multicolumn{4}{c}{13.81} \\ \hline
\multicolumn{5}{c}{(II) AUD} \\ \hline
\multicolumn{1}{l}{(e) HMM~\cite{ondel2016variational}} & \multicolumn{4}{c}{59.39} \\
\multicolumn{1}{l}{(f) SHMM~\cite{ondel2019bayesian}} & \multicolumn{4}{c}{55.69} \\
\multicolumn{1}{l}{(g) H-SHMM~\cite{yusuf2021hierarchical}} & \multicolumn{4}{c}{55.76} \\ \hline
\multicolumn{5}{c}{(III) Unpaired} \\ \hline
\multicolumn{1}{l}{\multirow{2}{*}{}} & \multicolumn{2}{c}{\emph{orc}} & \multicolumn{2}{c}{\emph{uns}} \\ \cline{2-5} 
\multicolumn{1}{l}{} & \emph{match} & \emph{nonmatch} & \emph{match} & \emph{nonmatch} \\ \hline
\multicolumn{1}{l}{(h) empirical-ODM~\cite{yeh2018unsupervised}} & 32.50 & 40.10 & 36.50 & 41.60 \\
\multicolumn{1}{l}{(i) proposed} & 28.74 & 31.81 & 31.03 & 36.71 \\
\thickhline
\end{tabular}
\label{tab:timit_whole}
\end{table}
Three types of previous methods are compared, and the results are shown in Table~\ref{tab:timit_whole}.
In (I), all supervised approaches are trained with labeled transcriptions on the standard TIMIT training set.
\emph{Supervised generator} (row (c)) and FS-RNN (row (d)) have additional access to the oracle boundary annotations.
In the \emph{supervised generator} (row (c)), we use the same architecture as the frame-wise generator and use the same WFST to decode the phone sequence, except trained by minimizing cross-entropy loss with frame-wise phone labels.
FS-RNN~\cite{ratajczak2017frame} (row (d)) is the previously proposed method, which uses a stacked frame-level and segment-level recurrent network.
In (II), three AUD models are compared.
Because AUD models only generate phone-like acoustic tokens, which cannot be used to calculate PER, we instead calculate `best-mapping PER' as followed steps: 
\begin{enumerate}[]
\item AUD models are first trained in an unsupervised manner to discover frame-wise acoustic tokens for each utterance, including training and testing utterances.
Following \cite{yusuf2021hierarchical}, the acoustic token truncation number is set to 100.
\item We use oracle frame-wise phone labels in the training set to get the best mapping from acoustic token to phone.
\item Utterances in the testing set are mapped into phone sequence and evaluated PER.
\end{enumerate}
‘Best-mapping PER’ has some advantages in the comparison because it uses the ground-truth information from the ground-truth frame-level phone sequences.
In (III), row (d) is the previously proposed baseline learning from unpaired speech and phone sequences, which is based on segmental empirical output distribution matching~\cite{yeh2018unsupervised}.

In the \emph{orc/match} case, the PER in row (i) achieved 28.74\%.
Although the performance is still far behind the strong supervised baselines in rows (a) (b) (d), it performs close to the \emph{supervised generator} (23.24\%) in row (c).
Both row (c) and \emph{orc/match} in row (i) can access the phone boundaries and the paired utterances and phone transcriptions.
The only difference is that the utterances and transcriptions are aligned in row (c) while not aligned in the \emph{orc/match} setting in row (i).
This is where the performance gap comes from.
All settings of the proposed approach perform better than the `best-mapping PER' of the three AUD models and perform consistently  better than the previous baseline Segmental Empirical-ODM in row (h).
Not to mention the prior work of Segmental Empirical-ODM needs a large batch size (up to 20000 training examples in a batch) to achieve a satisfactory performance, while the training process here is done with batch size as small as 100. 

\begin{figure}
\centering
\includegraphics[width=\linewidth]{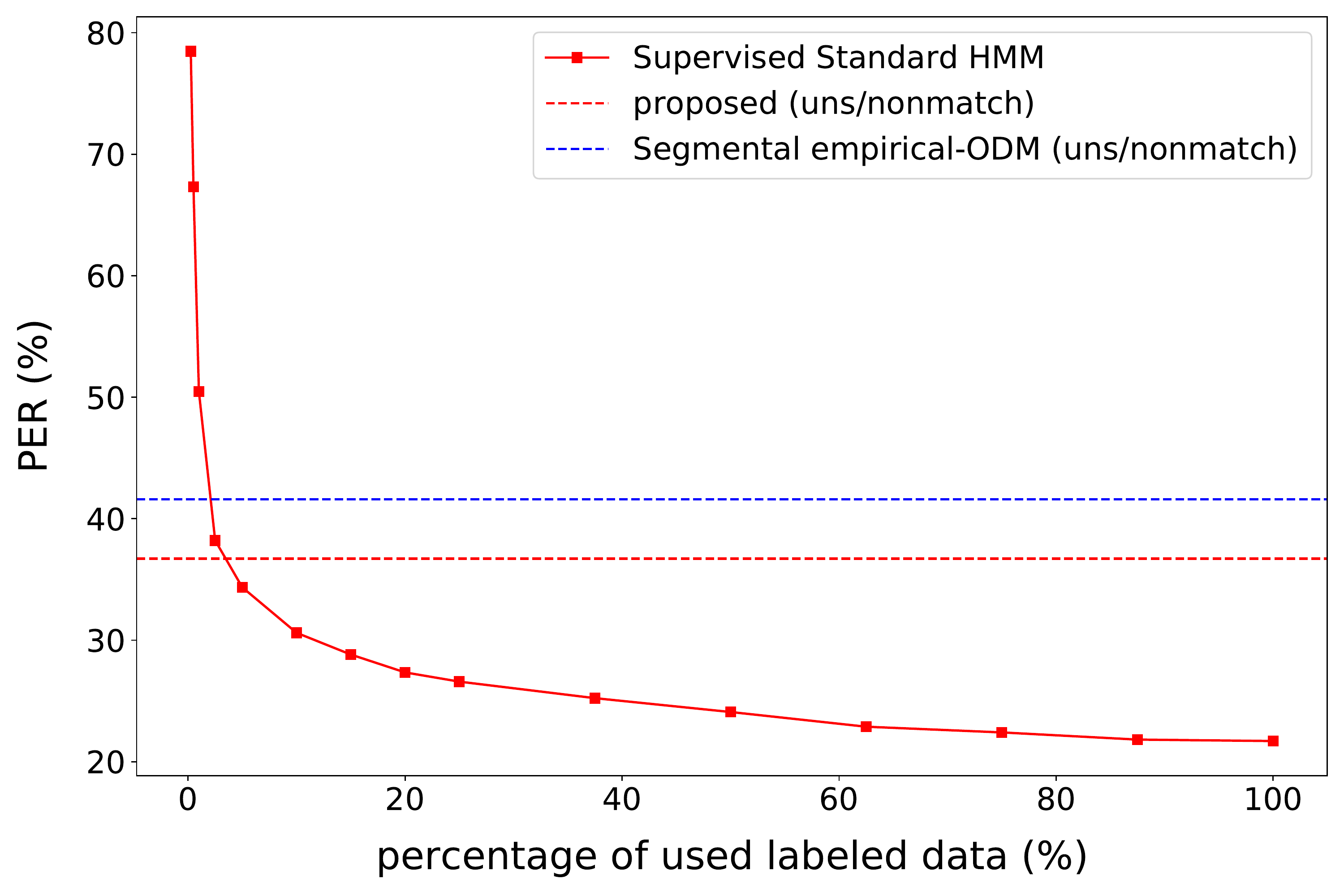}
\caption{Comparison of the proposed approaches to standard
supervised HMMs with varying quantity of labeled data.}
\label{fig:timit reduced label}
\end{figure}

Next, we want to determine the quantity of the labeled data required for the standard HMM to achieve comparable results to the proposed approach, illustrated in Fig~\ref{fig:timit reduced label}.
The red curve is for the standard HMMs trained with different portions of labeled data.
The red and blue dashed horizontal lines represent the proposed method and previous baseline Segmental empirical-ODM with \emph{uns}/\emph{nonmatch} setting.
We see the proposed method equals the situation when around 2.5\% to 5\% labeled data is used, while Segmental empirical-ODM is only comparable to use 1\% to 2.5\% labeled data.

\section{Discussion}
\subsection{Effectiveness of Self Re-training}\label{exp:Refinement}

\begin{table*}[]
\caption{
Performance of the framework over iterations.
\emph{Evaluation metric: PER in [\%].}
}
\centering
\begin{tabular}{clcccc}
\thickhline
\multicolumn{2}{c}{\multirow{2}{*}{}} & \multicolumn{2}{c}{\emph{orc}} & \multicolumn{2}{c}{\emph{uns}} \\ \cline{3-6} 
\multicolumn{2}{c}{} & \begin{tabular}[c]{@{}c@{}}\emph{match}\end{tabular} & \begin{tabular}[c]{@{}c@{}}\emph{nonmatch}\end{tabular} & \begin{tabular}[c]{@{}c@{}}\emph{match}\end{tabular} & \begin{tabular}[c]{@{}c@{}}\emph{nonmatch}\end{tabular} \\ \hline
\multicolumn{1}{c}{\multirow{2}{*}{iteration1}} & (a) GAN + WFST & 29.42 & \multicolumn{1}{c}{38.7} & 65.12 & 61.13 \\
\multicolumn{1}{c}{} & (b) GAN + WFST + HMM & 28.74 & \multicolumn{1}{c}{31.81} & 56.26 & 49.89 \\ \hline
\multicolumn{1}{c}{\multirow{2}{*}{iteration2}} & (c) GAN + WFST & \multicolumn{2}{c}{\multirow{2}{*}{-}} & 48.69 & 53.42 \\
\multicolumn{1}{c}{} & (d) GAN + WFST + HMM & \multicolumn{2}{c}{} & 33.96 & 41.08 \\ \hline
\multicolumn{1}{c}{\multirow{2}{*}{iteration3}} & (e) GAN + WFST & \multicolumn{2}{c}{\multirow{2}{*}{-}} & 43.33 & 52.31 \\
\multicolumn{1}{c}{} & (f) GAN + WFST + HMM & \multicolumn{2}{c}{} & 31.47 & 37.82 \\
\hline
\multicolumn{1}{c}{\multirow{2}{*}{iteration4}} & (g) GAN + WFST & \multicolumn{2}{c}{\multirow{2}{*}{-}} & 41.22 & 51.43 \\
\multicolumn{1}{c}{} & (h) GAN + WFST + HMM & \multicolumn{2}{c}{} & 31.03 & 36.71 \\
\thickhline
\end{tabular}
\label{tab:multi_performance}
\end{table*}

\begin{table*}[]
\caption{The performance improvement under different PER of the 1st iteration GAN training.
\emph{Evaluation metric: PER in [\%].}
}
\centering
\begin{tabular}{lcccccc}
\thickhline
iteration 1 GAN + WFST & 61.13 & 64.98 & 67.77 & 69.99 & 72.45 & 75.00 \\ 
iteration 2 GAN + WFST & 53.42 & 58.85 & 61.53 & 65.92 & 69.51 & 72.31 \\ \hline
improvement & 7.71 & 6.13 & 6.24 & 4.07 & 2.94 & 2.69 \\ 
\thickhline
\end{tabular}
\label{Tab:timit_diff_first}
\end{table*}

In this section, we investigate how our framework benefits from self re-training.
The results over different iterations are shown in Table \ref{tab:multi_performance}.
The rows (a)-(h) are for the iteration 1, 2, 3 and 4 of the framework, rows (a) (c) (e) (g) for GAN alone and decode with WFST decoder, while rows (b) (d) (f) (h) further use HMM self re-training. 
We can see the performance improves consistently after HMM self re-training at each iteration (rows (b) v.s. (a), (d) v.s. (c), (f) v.s. (e), (h) v.s. (g)). The improvement indicates the HMM re-training is beneficial even though the training transcription, which is the output of GAN, may be noisy.
Second, there is performance improvement after each iteration for either GAN alone (rows (a) (c) (e) (g)) or after self re-training (rows (b) (d) (f) (h)).
Because the only difference between iterations' training settings is the segmentation, we hypothesize that this improvement is due to the refinement of the phone boundaries over iterations. 

\begin{figure*}
\centering
\includegraphics[width=\linewidth]{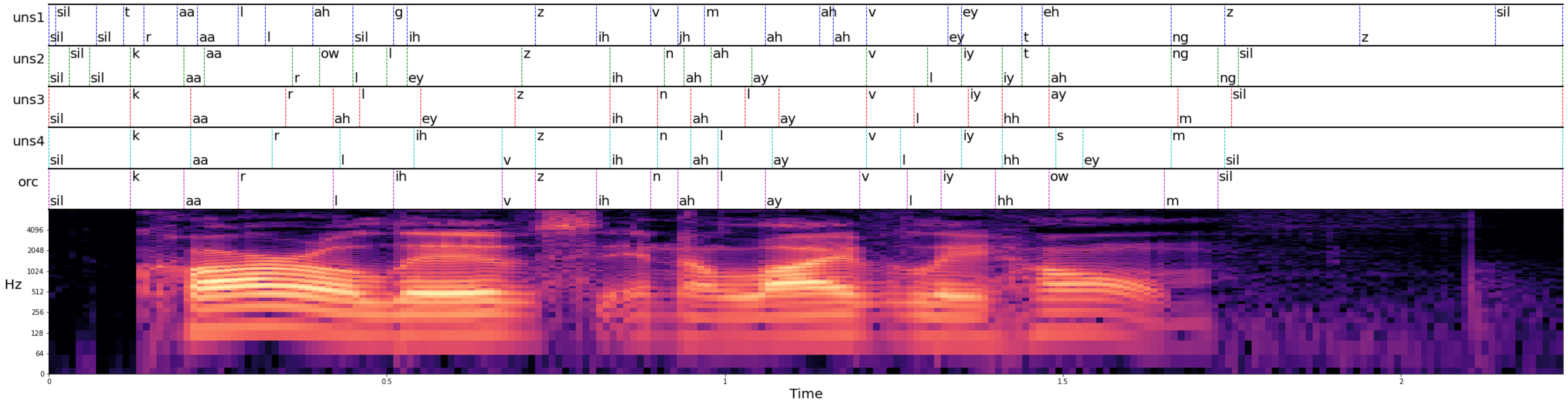}
\caption{The segmentation refinement of an example utterance over iterations under \emph{uns/match} setting. 
Specifically, the first row denotes the segmentation generated from GAS; 
the second row denotes the segmentation after the 1st iteration self re-training and forced alignment; 
the last row represents the oracle phone segmentation.
We can see the segmentation quality increases over the iterations in two different ways: 1. the generated segmentation is closer and closer to the oracle phone boundaries. 2. the redundant segmentation decreases. }
\label{fig:bnd_vis}
\end{figure*}

When evaluating the phone segmentation quality, the first thought is to use the F1-score, the harmonic mean of the precision and recall.
However, it is well-known that the F1-score is not suitable for segmentation because over-segmentation may give a very high recall leading to a high F1-score, even with a relatively low precision~\cite{michel2016blind}.
A naive periodic predictor, which predicts a phone boundary for every 40 ms, can still generate boundaries with precision 0.55, recall 0.99, and F1-score 0.71. 
We can see the high F1 score cannot reflect the poor quality of the predicted boundaries.
Therefore, this paper further adopts another better evaluation metric, R-value~\cite{rasanen2009improved}, which appropriately penalizes the over-segmentation phenomenon.
The R-value for the 40-ms periodic predictor is only 0.3, which gives a low score to the bad segmentation. 
The result is shown in Table~\ref{Tab:r-value}.
\begin{table}[]
\caption{phone segmentation qualities over iterations under \emph{uns/match} setting. Segmentation within 20-ms tolerance window is counted as correct segmentation. 
}
\centering
\begin{tabular}{lcc}
\thickhline
 & F1-score & R-value \\ \hline
40-ms periodic predictor & 0.71 & 0.3 \\ \hline
iteration1 & 0.76 & 0.72 \\ 
iteration2 & 0.79 & 0.76 \\ 
iteration3 & 0.8 & 0.77 \\ 
iteration4 & 0.81 & 0.78 \\ 
\thickhline
\end{tabular}
\label{Tab:r-value}
\end{table}
Both F1-score and R-value consistently increase and gradually converge over the iteration, which supports our hypothesis that self-training gives better and better segmentation over iterations.
Figure~\ref{fig:bnd_vis} also illustrates an example utterance of how the segmentation quality changes over the iterations.
Under \emph{uns} setting, because the phone boundaries become better and better over iterations, the training situation will be more and more similar to \emph{orc} setting, and finally get closer performance (Table \ref{tab:multi_performance}, \emph{orc} in (b) vs. \emph{uns} in (h)).

\subsection{Robustness of Self Re-training}
\label{subsec:stability}
In this section, we discuss how robust the self re-training is to the 1st iteration GAN training.
First, we run the GAN training nine times under \emph{uns/nonmatch} setting.
The mean and standard deviation of the performances are $67.88\pm4.27$.
Then we hand-pick several checkpoints of the 1st iteration GAN training that performs different PER.
The probing range of PER is roughly equally spaced between 61.13 (best PER) to 75 (approximately $mean+2*std$).
Each selected checkpoint is used to run the next pipeline iteration (self re-training + 2nd iteration GAN training).
The results are shown in Table~\ref{Tab:timit_diff_first}.
We observe that the self re-training can consistently get PER improvement even as the 1st iteration PER becomes worse (within the probing range).
This improvement becomes smaller as the PER of the 1st iteration GAN training becomes worse.

\subsection{Difference with the Concurrent Work: wav2vec-U}
This section highlights the main difference in the design between our work and the concurrent work wav2vec-U~\cite{baevski2021unsupervised}, which is based on our previous conference work~\cite{chen2019completely}. 
We categorize the difference into three points. 
\subsubsection{Feature} 
Our work uses hand-crafted features, MFCC.
Wav2vec-U selects the input feature from different layers of wave2vec 2.0~\cite{baevski2020wav2vec}, a self-supervised model.
The selecting criterion is the PER by training a linear model in a supervised manner.
\subsubsection{Unsupervised Segmentation}
Our work uses GAS~\cite{wang2017gate} to get the phone-level segmentation.
Wav2vec-U uses k-means to cluster the selected features, and the boundaries are drawn whenever the clustered index changes.

\subsubsection{Segmentation Refinement}
Our work uses HMM self re-training and forced alignment to refine the phone boundaries.
Wav2vec-U does not have explicit segmentation refinement.
While Wav2vec-U merges the neighboring segments containing the same predicted labels in each step of GAN training, this design can be viewed as refining the segmentation implicitly.

\section{Conclusion}
In this work, we take a step close to our vision `unsupervised ASR' by proposing a two-stage iterative framework to learn phone recognition from unpaired data. 
First, GAN training is shown able to learn the mapping between two domains: speech and phone sequence.
Then we categorize the generator into segment-wise and frame-wise generators, which are shown to be good at different settings.
Frame-wise generator performs well in \emph{uns} setting, while segment-wise generator, on the other hand, performs well in \emph{orc} setting.
Finally, self re-training is shown to improves the segmentation quality and boost the performance, and achieves 36.7\% under \emph{uns/nonmatch} setting over iterations.

\section{Acknowledgement}
We thank to National Center for High-performance Computing (NCHC) of National Applied Research Laboratories (NARLabs) in Taiwan for providing computational and storage resources.


%

\ifCLASSOPTIONcaptionsoff
  \newpage
\fi



%

\bibliography{mybib}

\begin{thebibliography}{10}
\providecommand{\url}[1]{#1}
\csname url@samestyle\endcsname
\providecommand{\newblock}{\relax}
\providecommand{\bibinfo}[2]{#2}
\providecommand{\BIBentrySTDinterwordspacing}{\spaceskip=0pt\relax}
\providecommand{\BIBentryALTinterwordstretchfactor}{4}
\providecommand{\BIBentryALTinterwordspacing}{\spaceskip=\fontdimen2\font plus
\BIBentryALTinterwordstretchfactor\fontdimen3\font minus
  \fontdimen4\font\relax}
\providecommand{\BIBforeignlanguage}[2]{{%
\expandafter\ifx\csname l@#1\endcsname\relax
\typeout{** WARNING: IEEEtran.bst: No hyphenation pattern has been}%
\typeout{** loaded for the language `#1'. Using the pattern for}%
\typeout{** the default language instead.}%
\else
\language=\csname l@#1\endcsname
\fi
#2}}
\providecommand{\BIBdecl}{\relax}
\BIBdecl

\bibitem{chorowski2015attention}
J.~K. Chorowski, D.~Bahdanau, D.~Serdyuk, K.~Cho, and Y.~Bengio,
  ``Attention-based models for speech recognition,'' in \emph{Advances in
  neural information processing systems}, 2015, pp. 577--585.

\bibitem{chiu2018state}
C.-C. Chiu, T.~N. Sainath, Y.~Wu, R.~Prabhavalkar, P.~Nguyen, Z.~Chen,
  A.~Kannan, R.~J. Weiss, K.~Rao, E.~Gonina \emph{et~al.}, ``State-of-the-art
  speech recognition with sequence-to-sequence models,'' in \emph{2018 IEEE
  International Conference on Acoustics, Speech and Signal Processing
  (ICASSP)}.\hskip 1em plus 0.5em minus 0.4em\relax IEEE, 2018, pp. 4774--4778.

\bibitem{gulati2020conformer}
A.~Gulati, J.~Qin, C.-C. Chiu, N.~Parmar, Y.~Zhang, J.~Yu, W.~Han, S.~Wang,
  Z.~Zhang, Y.~Wu \emph{et~al.}, ``Conformer: Convolution-augmented transformer
  for speech recognition,'' \emph{Proc. Interspeech 2020}, pp. 5036--5040,
  2020.

\bibitem{han2020contextnet}
W.~Han, Z.~Zhang, Y.~Zhang, J.~Yu, C.-C. Chiu, J.~Qin, A.~Gulati, R.~Pang, and
  Y.~Wu, ``Contextnet: Improving convolutional neural networks for automatic
  speech recognition with global context,'' \emph{Proc. Interspeech 2020}, pp.
  3610--3614, 2020.

\bibitem{artetxe2017unsupervised}
M.~Artetxe, G.~Labaka, E.~Agirre, and K.~Cho, ``Unsupervised neural machine
  translation,'' \emph{International Conference on Learning Representations},
  2018.

\bibitem{conneau2017word}
A.~Conneau, G.~Lample, M.~Ranzato, L.~Denoyer, and H.~J{\'e}gou, ``Word
  translation without parallel data,'' \emph{International Conference on
  Learning Representations}, 2018.

\bibitem{lample2017unsupervised}
G.~Lample, A.~Conneau, L.~Denoyer, and M.~Ranzato, ``Unsupervised machine
  translation using monolingual corpora only,'' \emph{International Conference
  on Learning Representations}, 2018.

\bibitem{goodfellow2014generative}
I.~Goodfellow, J.~Pouget-Abadie, M.~Mirza, B.~Xu, D.~Warde-Farley, S.~Ozair,
  A.~Courville, and Y.~Bengio, ``Generative adversarial nets,'' in
  \emph{Advances in neural information processing systems}, 2014, pp.
  2672--2680.

\bibitem{arjovsky2017wasserstein}
M.~Arjovsky, S.~Chintala, and L.~Bottou, ``Wasserstein gan,'' \emph{Proceedings
  of the 34th International Conference on Machine Learning}, pp. 214--223,
  2017.

\bibitem{gulrajani2017improved}
I.~Gulrajani, F.~Ahmed, M.~Arjovsky, V.~Dumoulin, and A.~Courville, ``Improved
  training of wasserstein gans,'' \emph{Proceedings of the 31st International
  Conference on Neural Information Processing Systems}, pp. 5769--5779, 2017.

\bibitem{wang2017gate}
Y.-H. Wang, C.-T. Chung, and H.-y. Lee, ``Gate activation signal analysis for
  gated recurrent neural networks and its correlation with phoneme
  boundaries,'' \emph{Proc. Interspeech 2017}, pp. 3822--3826, 2017.

\bibitem{michel2016blind}
P.~Michel, O.~R{\"a}s{\"a}nen, R.~Thiolliere, and E.~Dupoux, ``Blind phoneme
  segmentation with temporal prediction errors,'' \emph{Proceedings of ACL
  2017, Student Research Workshop}, pp. 62--68, 2017.

\bibitem{qiao2008unsupervised}
Y.~Qiao, N.~Shimomura, and N.~Minematsu, ``Unsupervised optimal phoneme
  segmentation: Objectives, algorithm and comparisons,'' in \emph{2008 IEEE
  International Conference on Acoustics, Speech and Signal Processing}.\hskip
  1em plus 0.5em minus 0.4em\relax IEEE, 2008, pp. 3989--3992.

\bibitem{kreuk2020self}
F.~Kreuk, J.~Keshet, and Y.~Adi, ``Self-supervised contrastive learning for
  unsupervised phoneme segmentation,'' \emph{Proc. Interspeech 2020}, pp.
  3700--3704, 2020.

\bibitem{franke2016phoneme}
J.~Franke, M.~Mueller, F.~Hamlaoui, S.~Stueker, and A.~Waibel, ``Phoneme
  boundary detection using deep bidirectional lstms,'' in \emph{Speech
  Communication; 12. ITG Symposium}.\hskip 1em plus 0.5em minus 0.4em\relax
  VDE, 2016, pp. 1--5.

\bibitem{rasanen2014basic}
O.~Rasanen, ``Basic cuts revisited: Temporal segmentation of speech into
  phone-like units with statistical learning at a pre-linguistic level,'' in
  \emph{Proceedings of the Annual Meeting of the Cognitive Science Society},
  vol.~36, no.~36, 2014.

\bibitem{kamper2020towards}
H.~Kamper and B.~van Niekerk, ``Towards unsupervised phone and word
  segmentation using self-supervised vector-quantized neural networks,''
  \emph{arXiv preprint arXiv:2012.07551}, 2020.

\bibitem{park2006unsupervised}
A.~Park and J.~R. Glass, ``Unsupervised word acquisition from speech using
  pattern discovery,'' in \emph{2006 IEEE International Conference on Acoustics
  Speech and Signal Processing Proceedings}, vol.~1.\hskip 1em plus 0.5em minus
  0.4em\relax IEEE, 2006, pp. I--I.

\bibitem{ten2007computational}
L.~ten Bosch and B.~Cranen, ``A computational model for unsupervised word
  discovery,'' 2007.

\bibitem{aimetti2009modelling}
G.~Aimetti, ``Modelling early language acquisition skills: Towards a general
  statistical learning mechanism,'' in \emph{Proceedings of the Student
  Research Workshop at EACL 2009}, 2009, pp. 1--9.

\bibitem{levin2013fixed}
K.~Levin, K.~Henry, A.~Jansen, and K.~Livescu, ``Fixed-dimensional acoustic
  embeddings of variable-length segments in low-resource settings,'' in
  \emph{2013 IEEE Workshop on Automatic Speech Recognition and
  Understanding}.\hskip 1em plus 0.5em minus 0.4em\relax IEEE, 2013, pp.
  410--415.

\bibitem{levin2015segmental}
K.~Levin, A.~Jansen, and B.~Van~Durme, ``Segmental acoustic indexing for zero
  resource keyword search,'' in \emph{2015 IEEE International Conference on
  Acoustics, Speech and Signal Processing (ICASSP)}.\hskip 1em plus 0.5em minus
  0.4em\relax IEEE, 2015, pp. 5828--5832.

\bibitem{bengio2014word}
S.~Bengio and G.~Heigold, ``Word embeddings for speech recognition,'' 2014.

\bibitem{chen2015query}
G.~Chen, C.~Parada, and T.~N. Sainath, ``Query-by-example keyword spotting
  using long short-term memory networks,'' in \emph{2015 IEEE International
  Conference on Acoustics, Speech and Signal Processing (ICASSP)}.\hskip 1em
  plus 0.5em minus 0.4em\relax IEEE, 2015, pp. 5236--5240.

\bibitem{kamper2016deep}
H.~Kamper, W.~Wang, and K.~Livescu, ``Deep convolutional acoustic word
  embeddings using word-pair side information,'' in \emph{2016 IEEE
  International Conference on Acoustics, Speech and Signal Processing
  (ICASSP)}.\hskip 1em plus 0.5em minus 0.4em\relax IEEE, 2016, pp. 4950--4954.

\bibitem{he2016multi}
W.~He, W.~Wang, and K.~Livescu, ``Multi-view recurrent neural acoustic word
  embeddings,'' \emph{arXiv preprint arXiv:1611.04496}, 2016.

\bibitem{settle2017query}
S.~Settle, K.~Levin, H.~Kamper, and K.~Livescu, ``Query-by-example search with
  discriminative neural acoustic word embeddings,'' \emph{Proc. Interspeech
  2017}, pp. 2874--2878, 2017.

\bibitem{maas2012word}
A.~L. Maas, S.~D. Miller, T.~M. O’neil, A.~Y. Ng, and P.~Nguyen, ``Word-level
  acoustic modeling with convolutional vector regression,'' in \emph{Proc. ICML
  Workshop Representation Learn}, 2012.

\bibitem{chung2018speech2vec}
Y.-A. Chung and J.~Glass, ``Speech2vec: A sequence-to-sequence framework for
  learning word embeddings from speech,'' \emph{Proc. Interspeech 2018}, pp.
  811--815, 2018.

\bibitem{holzenberger2018learning}
N.~Holzenberger, M.~Du, J.~Karadayi, R.~Riad, and E.~Dupoux, ``Learning word
  embeddings: Unsupervised methods for fixed-size representations of
  variable-length speech segments,'' 2018, pp. 2683--2687.

\bibitem{kamper2019truly}
H.~Kamper, ``Truly unsupervised acoustic word embeddings using weak top-down
  constraints in encoder-decoder models,'' in \emph{ICASSP 2019-2019 IEEE
  International Conference on Acoustics, Speech and Signal Processing
  (ICASSP)}.\hskip 1em plus 0.5em minus 0.4em\relax IEEE, 2019, pp. 6535--3539.

\bibitem{settle2016discriminative}
S.~Settle and K.~Livescu, ``Discriminative acoustic word embeddings: Tecurrent
  neural network-based approaches,'' in \emph{2016 IEEE Spoken Language
  Technology Workshop (SLT)}.\hskip 1em plus 0.5em minus 0.4em\relax IEEE,
  2016, pp. 503--510.

\bibitem{chung2016audio}
Y.-A. Chung, C.-C. Wu, C.-H. Shen, H.-Y. Lee, and L.-S. Lee, ``Audio word2vec:
  Unsupervised learning of audio segment representations using
  sequence-to-sequence autoencoder,'' \emph{Interspeech 2016}, pp. 765--769,
  2016.

\bibitem{dunbar2019zero}
E.~Dunbar, R.~Algayres, J.~Karadayi, M.~Bernard, J.~Benjumea, X.-N. Cao,
  L.~Miskic, C.~Dugrain, L.~Ondel, A.~W. Black \emph{et~al.}, ``The zero
  resource speech challenge 2019: Tts without t,'' \emph{Proc. Interspeech
  2019}, pp. 1088--1092, 2019.

\bibitem{dunbar2020zero}
E.~Dunbar, J.~Karadayi, M.~Bernard, X.-N. Cao, R.~Algayres, L.~Ondel,
  L.~Besacier, S.~Sakti, and E.~Dupoux, ``The zero resource speech challenge
  2020: Discovering discrete subword and word units,'' \emph{Proc. Interspeech
  2020}, pp. 4831--4835, 2020.

\bibitem{chen2020unsupervised}
M.~Chen and T.~Hain, ``Unsupervised acoustic unit representation learning for
  voice conversion using wavenet auto-encoders,'' \emph{Proc. Interspeech
  2020}, pp. 4866--4870, 2020.

\bibitem{baevski2019vq}
A.~Baevski, S.~Schneider, and M.~Auli, ``vq-wav2vec: Self-supervised learning
  of discrete speech representations,'' \emph{International Conference on
  Learning Representations}, 2019.

\bibitem{eloff2019unsupervised}
R.~Eloff, A.~Nortje, B.~van Niekerk, A.~Govender, L.~Nortje, A.~Pretorius,
  E.~Van~Biljon, E.~van~der Westhuizen, L.~van Staden, and H.~Kamper,
  ``Unsupervised acoustic unit discovery for speech synthesis using discrete
  latent-variable neural networks,'' \emph{Proc. Interspeech 2019}, pp.
  1103--1107, 2019.

\bibitem{chorowski2019unsupervised}
J.~Chorowski, R.~J. Weiss, S.~Bengio, and A.~van~den Oord, ``Unsupervised
  speech representation learning using wavenet autoencoders,'' \emph{IEEE/ACM
  transactions on audio, speech, and language processing}, vol.~27, no.~12, pp.
  2041--2053, 2019.

\bibitem{lee2012nonparametric}
C.-y. Lee and J.~Glass, ``A nonparametric bayesian approach to acoustic model
  discovery,'' in \emph{Proceedings of the 50th Annual Meeting of the
  Association for Computational Linguistics (Volume 1: Long Papers)}, 2012, pp.
  40--49.

\bibitem{ondel2016variational}
L.~Ondel, L.~Burget, and J.~{\v{C}}ernock{\`y}, ``Variational inference for
  acoustic unit discovery,'' \emph{Procedia Computer Science}, vol.~81, pp.
  80--86, 2016.

\bibitem{chen2017multilingual}
H.~Chen, C.-C. Leung, L.~Xie, B.~Ma, and H.~Li, ``Multilingual bottle-neck
  feature learning from untranscribed speech,'' in \emph{2017 IEEE Automatic
  Speech Recognition and Understanding Workshop (ASRU)}.\hskip 1em plus 0.5em
  minus 0.4em\relax IEEE, 2017, pp. 727--733.

\bibitem{ondel2019bayesian}
L.~Ondel, H.~K. Vydana, L.~Burget, and J.~{\v{C}}ernock{\`y}, ``Bayesian
  subspace hidden markov model for acoustic unit discovery,'' \emph{Proc.
  Interspeech 2019}, pp. 261--265, 2019.

\bibitem{yusuf2021hierarchical}
B.~Yusuf, L.~Ondel, L.~Burget, J.~{\v{C}}ernock{\`y}, and M.~Sara{\c{c}}lar,
  ``A hierarchical subspace model for language-attuned acoustic unit
  discovery,'' in \emph{ICASSP 2021-2021 IEEE International Conference on
  Acoustics, Speech and Signal Processing (ICASSP)}.\hskip 1em plus 0.5em minus
  0.4em\relax IEEE, 2021, pp. 3710--3714.

\bibitem{bansal2017towards}
S.~Bansal, H.~Kamper, A.~Lopez, and S.~Goldwater, ``Towards speech-to-text
  translation without speech recognition,'' \emph{EACL 2017}, p. 474, 2017.

\bibitem{liu2018completely}
D.-R. Liu, K.-Y. Chen, H.-Y. Lee, and L.-s. Lee, ``Completely unsupervised
  phoneme recognition by adversarially learning mapping relationships from
  audio embeddings,'' \emph{Proc. Interspeech 2018}, pp. 3748--3752, 2018.

\bibitem{liu2017unsupervised}
Y.~Liu, J.~Chen, and L.~Deng, ``Unsupervised sequence classification using
  sequential output statistics,'' in \emph{Advances in Neural Information
  Processing Systems}, 2017, pp. 3550--3559.

\bibitem{yeh2018unsupervised}
C.-K. Yeh, J.~Chen, C.~Yu, and D.~Yu, ``Unsupervised speech recognition via
  segmental empirical output distribution matching,'' \emph{International
  Conference on Learning Representations}, 2018.

\bibitem{chen2019completely}
K.-Y. Chen, C.-P. Tsai, D.-R. Liu, H.-Y. Lee, and L.-s. Lee, ``Completely
  unsupervised speech recognition by a generative adversarial network
  harmonized with iteratively refined hidden markov models,'' \emph{Proc.
  Interspeech 2019}, pp. 1856--1860, 2019.

\bibitem{chung2018unsupervised}
Y.-A. Chung, W.-H. Weng, S.~Tong, and J.~Glass, ``Unsupervised cross-modal
  alignment of speech and text embedding spaces,'' \emph{Advances in Neural
  Information Processing Systems}, vol.~31, pp. 7354--7364, 2018.

\bibitem{chung2019towards}
------, ``Towards unsupervised speech-to-text translation,'' in \emph{ICASSP
  2019-2019 IEEE International Conference on Acoustics, Speech and Signal
  Processing (ICASSP)}, 2019, pp. 7170--7174.

\bibitem{jang2016categorical}
E.~Jang, S.~Gu, and B.~Poole, ``Categorical reparameterization with
  gumbel-softmax,'' \emph{arXiv preprint arXiv:1611.01144}, 2016.

\bibitem{maddison2016concrete}
C.~J. Maddison, A.~Mnih, and Y.~W. Teh, ``The concrete distribution: A
  continuous relaxation of discrete random variables,'' \emph{Proceedings of
  the international conference on learning Representations}, 2017.

\bibitem{gumbel1954statistical}
\BIBentryALTinterwordspacing
E.~Gumbel, \emph{Statistical Theory of Extreme Values and Some Practical
  Applications: A Series of Lectures}, ser. Applied mathematics series.\hskip
  1em plus 0.5em minus 0.4em\relax U.S. Government Printing Office, 1954.
  [Online]. Available: \url{https://books.google.com.tw/books?id=SNpJAAAAMAAJ}
\BIBentrySTDinterwordspacing

\bibitem{wang2017tacotron}
Y.~Wang, R.~Skerry-Ryan, D.~Stanton, Y.~Wu, R.~J. Weiss, N.~Jaitly, Z.~Yang,
  Y.~Xiao, Z.~Chen, S.~Bengio \emph{et~al.}, ``Tacotron: Towards end-to-end
  speech synthesis,'' \emph{Proc. Interspeech 2017}, pp. 4006--4010, 2017.

\bibitem{lee2017fully}
J.~Lee, K.~Cho, and T.~Hofmann, ``Fully character-level neural machine
  translation without explicit segmentation,'' \emph{Transactions of the
  Association for Computational Linguistics}, vol.~5, pp. 365--378, 2017.

\bibitem{liu2019variance}
L.~Liu, H.~Jiang, P.~He, W.~Chen, X.~Liu, J.~Gao, and J.~Han, ``On the variance
  of the adaptive learning rate and beyond,'' \emph{International Conference on
  Learning Representations}, 2019.

\bibitem{vaswani2017attention}
A.~Vaswani, N.~Shazeer, N.~Parmar, J.~Uszkoreit, L.~Jones, A.~N. Gomez,
  {\L}.~Kaiser, and I.~Polosukhin, ``Attention is all you need,'' in
  \emph{Advances in neural information processing systems}, 2017, pp.
  5998--6008.

\bibitem{Povey_ASRU2011}
D.~Povey, A.~Ghoshal, G.~Boulianne, L.~Burget, O.~Glembek, N.~Goel,
  M.~Hannemann, P.~Motlicek, Y.~Qian, P.~Schwarz, J.~Silovsky, G.~Stemmer, and
  K.~Vesely, ``The kaldi speech recognition toolkit,'' in \emph{IEEE 2011
  Workshop on Automatic Speech Recognition and Understanding}.\hskip 1em plus
  0.5em minus 0.4em\relax IEEE Signal Processing Society, Dec. 2011, iEEE
  Catalog No.: CFP11SRW-USB.

\bibitem{ratajczak2017frame}
M.~Ratajczak, S.~Tschiatschek, and F.~Pernkopf, ``Frame and segment level
  recurrent neural networks for phone classification.'' in \emph{Interspeech},
  2017, pp. 1318--1322.

\bibitem{rasanen2009improved}
O.~J. R{\"a}s{\"a}nen, U.~K. Laine, and T.~Altosaar, ``An improved speech
  segmentation quality measure: the r-value,'' in \emph{Tenth Annual Conference
  of the International Speech Communication Association}, 2009.

\bibitem{baevski2021unsupervised}
A.~Baevski, W.-N. Hsu, A.~Conneau, and M.~Auli, ``Unsupervised speech
  recognition,'' \emph{arXiv preprint arXiv:2105.11084}, 2021.

\bibitem{baevski2020wav2vec}
A.~Baevski, H.~Zhou, A.~Mohamed, and M.~Auli, ``wav2vec 2.0: A framework for
  self-supervised learning of speech representations,'' \emph{Advances in
  Neural Information Processing Systems}, vol.~33, 2020.

\end{thebibliography}

\bibliographystyle{IEEEtran}

%


\begin{IEEEbiography}
[{\includegraphics[width=1in,height=1.25in,clip,keepaspectratio]{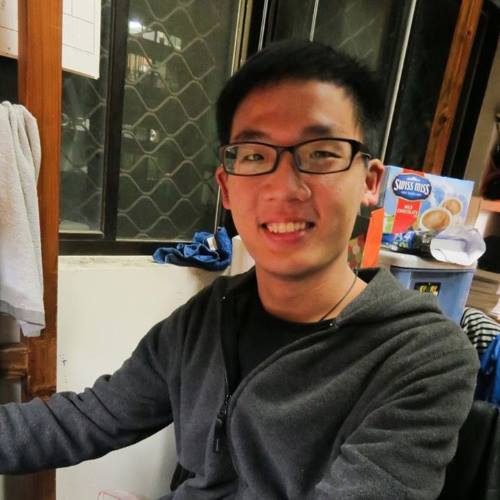} }]
{Da-rong Liu}
Da-rong Liu, received the Bachelor degree from National Taiwan University (NTU) in 2016, and is now a P.h. D. student at the Graduate Institute of Communication Engineering (GICE) at National Taiwan University. He mainly works on unsupervised learning, speech recognition and speech generation.
\end{IEEEbiography}

\begin{IEEEbiography}
[{\includegraphics[width=1in,height=1.25in,clip,keepaspectratio]{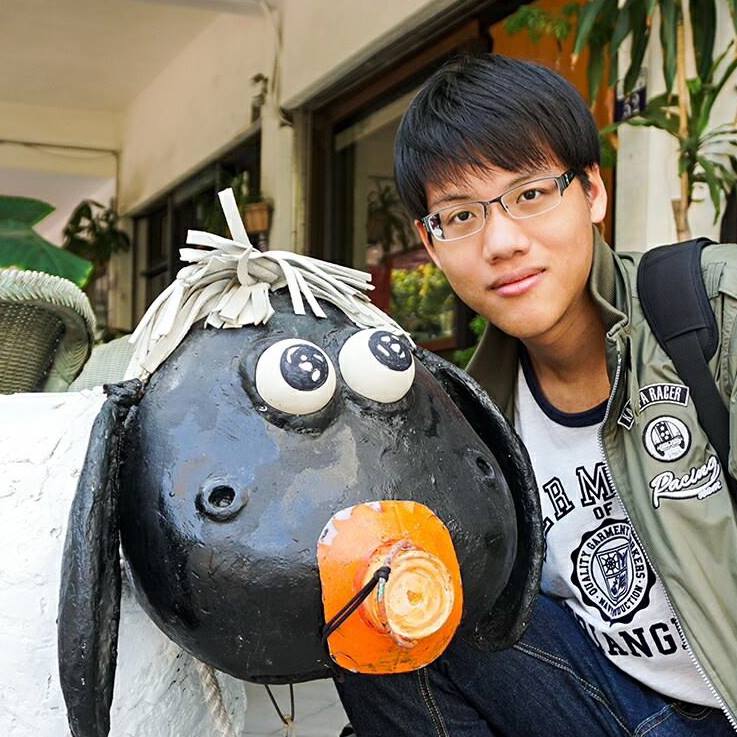} }]
{Po-chun Hsu}
received the B.S. degree from National Taiwan University (NTU) in 2018 and is now a P.h. D. student at the Graduate Institute of Communication Engineering (GICE) at NTU. His research focuses on speech synthesis, including text-to-speech (TTS), voice conversion (VC), and neural vocoder.
\end{IEEEbiography}

\begin{IEEEbiography}
[{\includegraphics[width=1in,height=1.25in,clip,keepaspectratio]{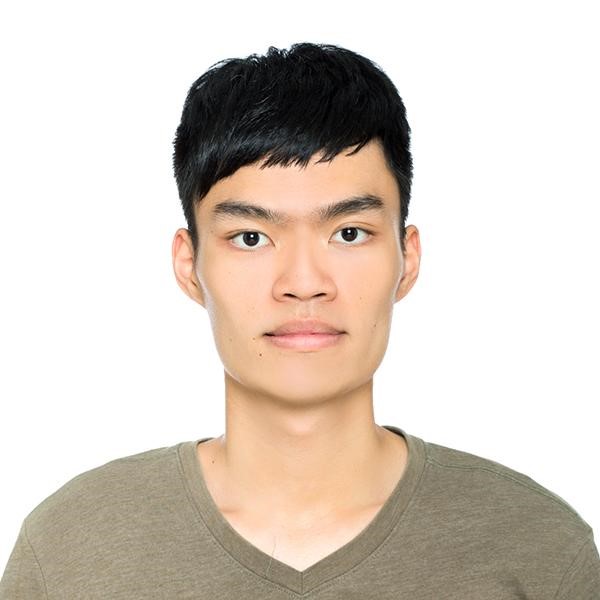} }]
{Yi-chen Chen}
Yi-chen Chen, received the Bachelor degree from National Taiwan University (NTU) in 2017, and is now a P.h. D. student at the Graduate Institute of Communication Engineering (GICE) at National Taiwan University, working on self-supervised/semi-supervised/transfer learning and speech processing.
\end{IEEEbiography}

\begin{IEEEbiography}
[{\includegraphics[width=1in,height=1.25in,clip,keepaspectratio]{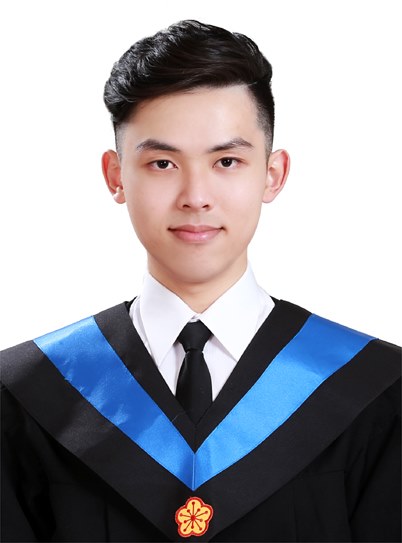} }]
{Sung-Feng Huang}
Sung-Feng Huang, received the Bachelor degree from National Taiwan University (NTU) in 2017, and is now a P.h. D. student at the Graduate Institute of Communication Engineering (GICE) at National Taiwan University. He mainly works on learning representations unsupervisedly, speech recognition, spoken term detection, speech separation, meta learning and machine learning techniques.
\end{IEEEbiography}

\begin{IEEEbiography}
[{\includegraphics[width=1in,height=1.25in,clip,keepaspectratio]{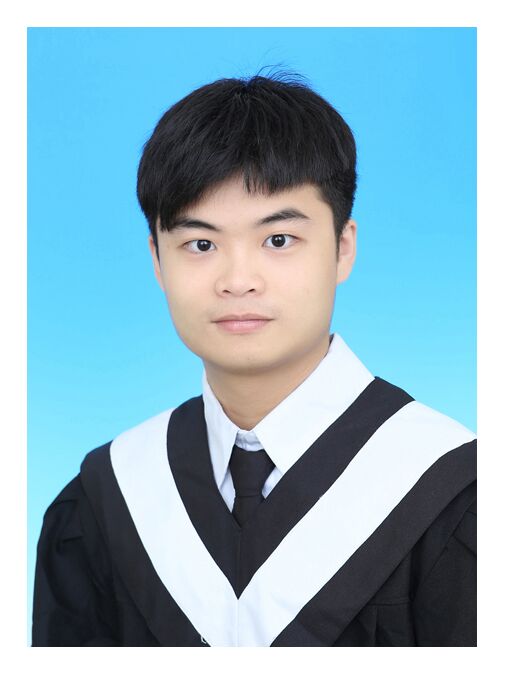} }]
{Shun-Po Chuang}
joined the Speech Processing and Machine Learning Laboratory in 2016.
He is currently working toward the Ph.D. degree at National Taiwan
University, Taipei, Taiwan. His research focuses on speech recognition, speech-to-text translation and 
code-switching language modeling.
\end{IEEEbiography}

\begin{IEEEbiography}
[{\includegraphics[width=1in,height=1.25in,clip,keepaspectratio]{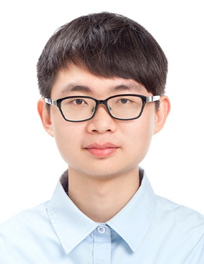} }]
{Da-yi Wu}
Da-Yi Wu received the B.S and M.S degrees from National Taiwan University, Taipei, Taiwan, in 2014 and 2018, respectively. He is currently a research assistant for NTU Speech Processing Laboratory. His research focuses on voice conversion, singing generation, and machine learning.
\end{IEEEbiography}

\begin{IEEEbiography}
[{\includegraphics[width=1in,height=1.25in,clip,keepaspectratio]{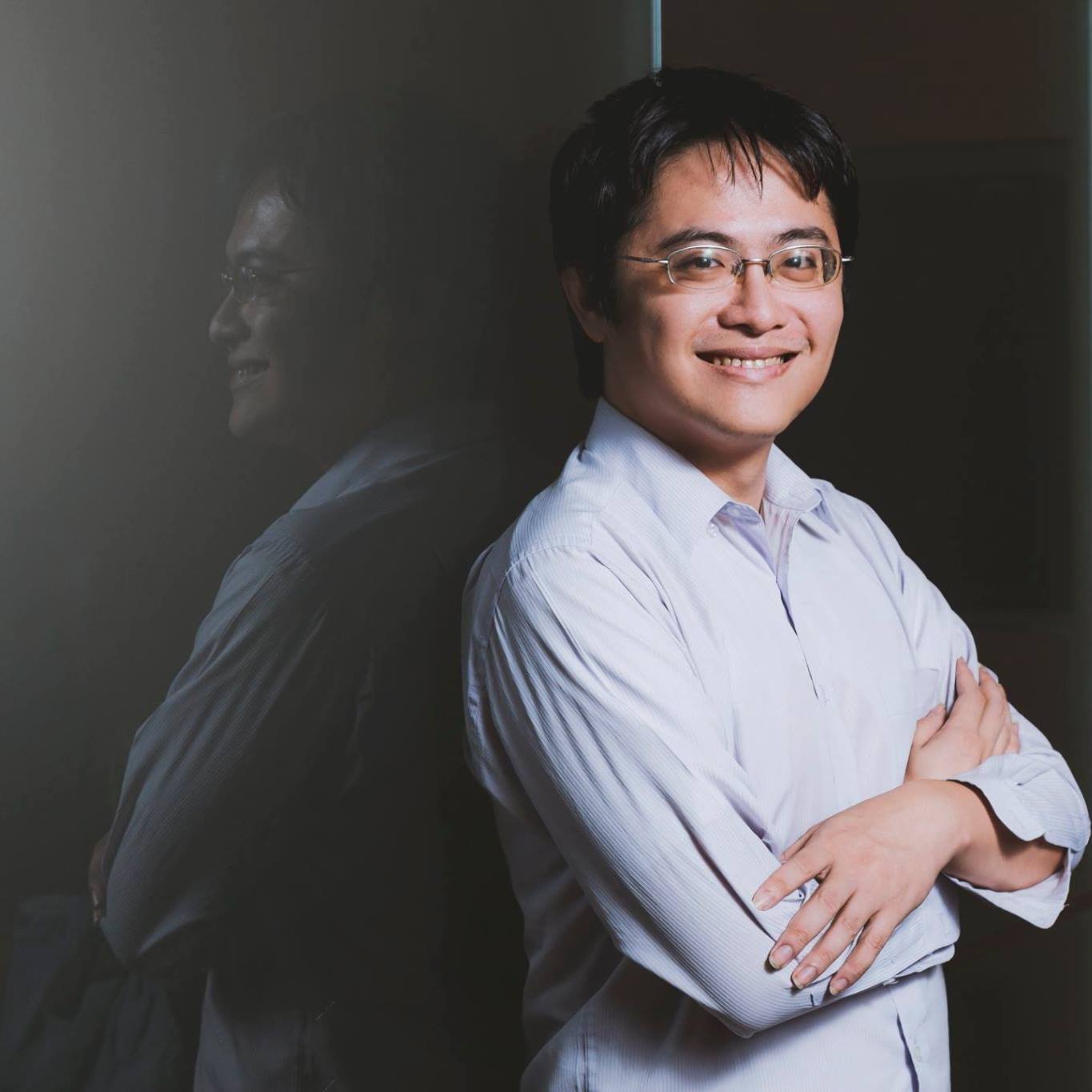} }]
{Hung-yi Lee}
received the M.S. and Ph.D. degrees from National Taiwan University, Taipei,
Taiwan, in 2010 and 2012, respectively. From September 2012 to August 2013, he
was a Postdoctoral Fellow with Research Center for Information Technology
Innovation, Academia Sinica. From September 2013 to July 2014, he was a
Visiting Scientist with the Spoken Language Systems Group, MIT Computer Science
and Artificial Intelligence Laboratory. He is currently an Associate Professor
with the Department of Electrical Engineering, National Taiwan University,
Taipei, Taiwan with a joint appointment to the Department of Computer Science and
Information Engineering. His research focuses on spoken
language understanding, speech recognition, and machine learning.

\end{IEEEbiography}

\end{document}